\documentclass[sigconf]{acmart}
\usepackage[table]{xcolor} 
\definecolor{LightGreen}{rgb}{0.85,1,0.85} 
\definecolor{LightRed}{rgb}{1,0.85,0.85} 

\usepackage{adjustbox} 

\usepackage{hyperref}
\usepackage{makecell}
\usepackage{booktabs}      
\usepackage{array}         
\usepackage{makecell}      
\usepackage{adjustbox}     
\usepackage[table]{xcolor} 

\newcolumntype{Y}[1]{>{\centering\arraybackslash}p{#1}}

\definecolor{LightGreen}{RGB}{219, 244, 223}
\definecolor{LightRed}{RGB}{245, 220, 220}

\renewcommand\footnotetextcopyrightpermission[1]{} 
\AtBeginDocument{%
  }

\setcopyright{acmlicensed}
\copyrightyear{2018}
\acmYear{2018}
\acmDOI{XXXXXXX.XXXXXXX}
\acmConference[Conference acronym 'XX]{Make sure to enter the correct
  conference title from your rights confirmation email}{June 03--05,
  2018}{Woodstock, NY}
\acmISBN{978-1-4503-XXXX-X/2018/06}

\begin{document}

\title{Attention Distance: A Novel Metric for Directed Fuzzing with Large Language Models}
\author{%
Wang Bin$^{1}$,
Ao Yang$^{1}$,
Kedan Li$^{2}$,
Aofan Liu$^{1}$,
Hui Li$^{1,*}$,
Guibo Luo$^{1,*}$,
Weixiang Huang$^{3}$,
Yan Zhuang$^{3}$%
}

\affiliation{%
  \institution{$^{1}$Guangdong Provincial Key Laboratory of Ultra High Definition Immersive Media Technology, Shenzhen Graduate School, Peking University \quad
               $^{2}$University of Illinois at Urbana-Champaign \quad
               $^{3}$China Mobile Internet CO}
  \country{}
}

\email{2201111747@stu.pku.edu.cn,
       jarvisya@stu.pku.edu.cn,
       kedanli2@illinois.edu,
       SWE2009510@xmu.edu.my}
\email{lih64@pkusz.edu.cn,
       luogb@pku.edu.cn,
       huangweixiang@cmic.chinamobile.com,
       zhuangyan@cmic.chinamopile.com}

\renewcommand{\shortauthors}{Trovato et al.}

\begin{abstract}

In the domain of software security testing, Directed Grey-Box Fuzzing (DGF) has garnered widespread attention for its efficient target localization and excellent detection performance. However, existing approaches measure only the physical distance between seed execution paths and target locations, overlooking logical relationships among code segments. This omission can yield redundant or misleading guidance in complex binaries, weakening DGF’s real-world effectiveness. To address this, we introduce \textbf{attention distance}, a novel metric that leverages a large language model’s contextual analysis to compute attention scores between code elements and reveal their intrinsic connections. Under the same AFLGo configuration—without altering any fuzzing components other than the distance metric—replacing physical distances with attention distances across 38 real vulnerability reproduction experiments delivers a \textbf{3.43×} average increase in testing efficiency over the traditional method. Compared to state-of-the-art directed fuzzers DAFL and WindRanger, our approach achieves \textbf{2.89×} and \textbf{7.13×} improvements, respectively. To further validate the generalizability of attention distance, we integrate it into DAFL and WindRanger, where it also consistently enhances their original performance. All related code and datasets are publicly available at https://anonymous.4open.science/r/Attention\_Distance-4650.

\end{abstract}



\received{20 February 2007}
\received[revised]{12 March 2009}
\received[accepted]{5 June 2009}

\maketitle

\section{Introduction}

Fuzz testing, a highly efficient automated software testing technique, involves injecting a large volume of random or semi-random data into a program to uncover potential errors and vulnerabilities. Directed fuzzing, in contrast to traditional fuzzing, focuses more on in-depth testing of specific error-prone areas within the software, rather than solely aiming for code coverage. This approach has shown significant advantages in scenarios such as vulnerability reproduction \cite{bohmeDirectedGreyboxFuzzing2017,chenHawkeyeDesiredDirected2018a,zongFuzzGuardFilteringOut2020,xuanCrashReproductionTest2015,wang2023syztrust,huang2024everything}.

Current research in directed fuzzing primarily focuses on the precise selection and optimization of input seeds \cite{shahMC2RigorousEfficient2022,zongFuzzGuardFilteringOut2020}, innovative distance measurement mechanisms \cite{chenHawkeyeDesiredDirected2018a,duWindRangerDirectedGreybox2022}, and optimization of static analysis \cite{huangBEACONDirectedGreybox2022,srivastavaOneFuzzDoesn2022,kimDAFLDirectedGreybox2023} to enhance the efficiency and accuracy of the testing process. By refining input seed selection, the testing effort can be more focused on code areas closely related to the target vulnerabilities, significantly reducing ineffective and redundant test attempts. Moreover, employing novel distance measurement mechanisms allows for accurate assessment of the proximity between input seeds and the target locations, optimizing testing paths. Additionally, static analysis of target code can effectively prune or adjust testing paths, further enhancing the performance of directed fuzzing.

\begin{figure}[h]
    \captionsetup{aboveskip=0.5pt, belowskip=0.5pt}
    \centering
    \includegraphics[width=0.75\columnwidth]{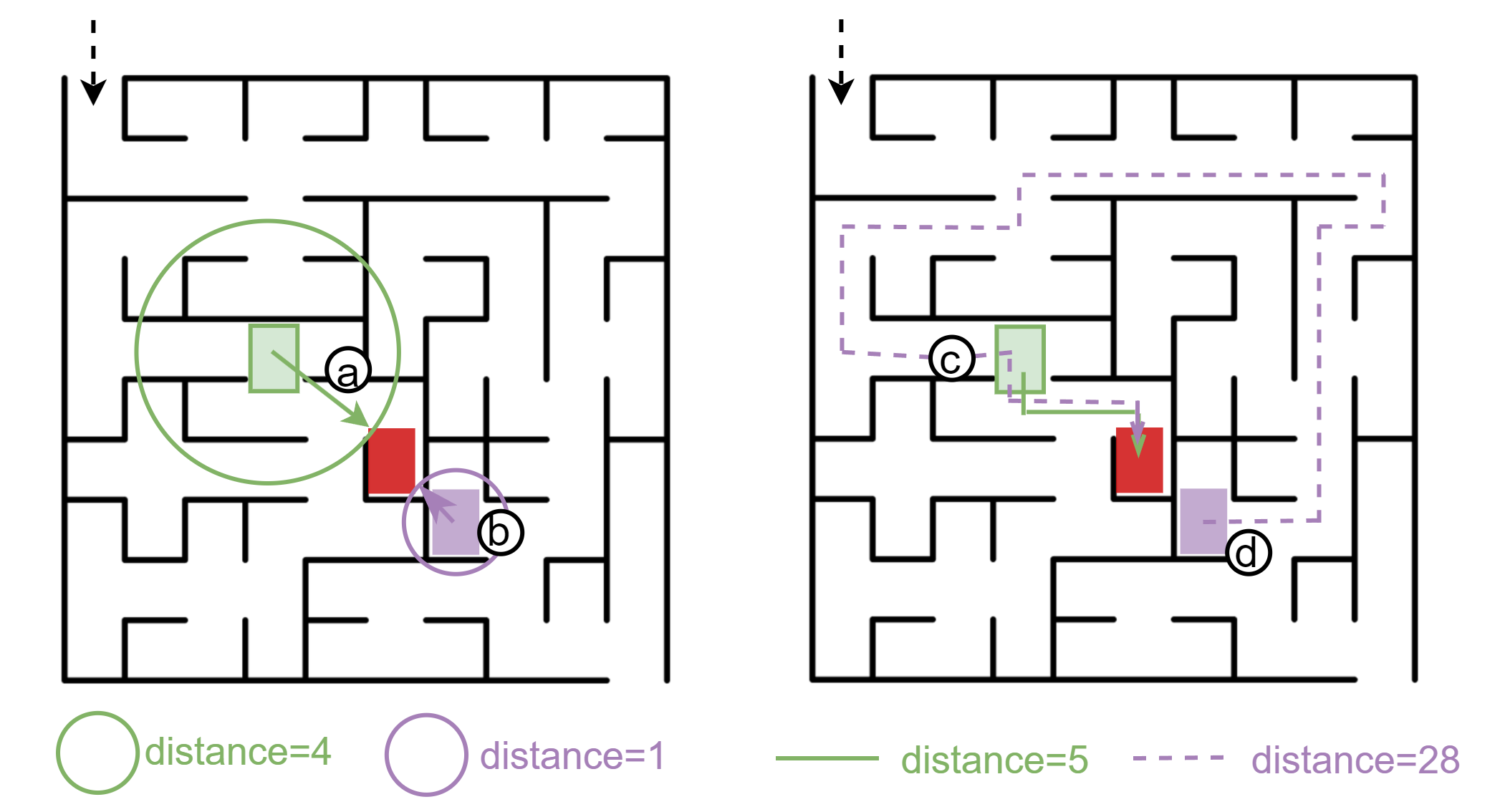}
    \caption{Maze Effect of distance calculation. Knowing the target and distance still cannot guarantee that a shorter distance means a closer exploration path. The red square represents the target point, and the green and purple squares represent two starting points with different distances.} 
    \label{fig:migong}
\end{figure}

The progression of directed fuzzing technology heavily relies on the distance-based strategy introduced by AFLGo\cite{bohmeDirectedGreyboxFuzzing2017}. This method enhances testing direction by assessing the distance between input seeds and target code segments, subsequently determining seed priority. While instrumental in clarifying test directions and priority settings, its primary limitation is its sole reliance on the structural information of code—that is, it only discerns the direct distance between the current position and the target without awareness of the actual route. This neglect of the logical relationships between code structures can lead to the erroneous triggering of irrelevant branches, resulting in incorrect priority assignment. This limitation mirrors the maze effect, as depicted in Figure \ref{fig:migong}. Although the evaluation method on the left side of the figure can measure the proximity of structural distances, it lacks the support of more comprehensive information. Consequently, even though testers are aware of endpoint locations and distances, they may still struggle to ensure the adoption of the most direct and effective testing pathways, occasionally misjudging path \textbf{b} to be closer to the target than path \textbf{a}. However, in practice, the paths that effectively reach the endpoint might be \textbf{c} and \textbf{d}, with path \textbf{c} presenting significantly less difficulty in reaching the target compared to path \textbf{d}. The left image in Figure \ref{fig:migong} resembles the current guidance based on distance, which only indicates straight-line distances. However, proximity does not necessarily simplify the challenge of triggering vulnerabilities. Certain vulnerabilities require specific preconditions or complex branch decisions to be triggered, which are not effectively reflected by mere distance metrics.

The advancement of Directed Grey-Box Fuzzing (DGF) owes much to the distance-based prioritization strategy introduced by AFLGo \cite{bohmeDirectedGreyboxFuzzing2017}. By computing the structural distance between each input seed’s execution path and a target code region, AFLGo ranks seeds to guide testing toward the vulnerability. However, this approach is fundamentally limited by its exclusive reliance on control-flow structure: it measures only the shortest static distance to the target, without any awareness of the actual execution routes or the logical dependencies among branches. As a result, the fuzzer may be drawn down irrelevant paths, misallocating priority to seeds that appear “closer” on paper but diverge in practice—an effect akin to navigating a maze (Figure \ref{fig:migong}). In that illustration, path \textbf{b} seems shorter than \textbf{a}, yet the true execution corridors \textbf{c} and \textbf{d} reveal that \textbf{c} actually offers a much simpler route to the goal than \textbf{d}. More generally, structural adjacency does not guarantee ease of reachability: many vulnerabilities require specific preconditions or branch-decision sequences that pure distance metrics cannot capture.

The advancements in Large Language Models (LLMs) within the field of software engineering have unlocked unprecedented possibilities. Leveraging deep learning technologies, LLMs' profound understanding and generative capabilities regarding programming languages have been proven to effectively assist developers in patching vulnerabilities \cite{wangEnhancingLargeLanguage2023,heLargeLanguageModels2023,de-fitero-dominguezEnhancedAutomatedCode2024}, translating code languages \cite{zhengCodeGeeXPretrainedModel2023}, and conducting in-depth analysis of complex codebases. Through meticulous tuning, LLMs have demonstrated exceptional abilities in code analysis \cite{wanImprovingAutomaticSource2018}, bug detection \cite{mathewsLLbezpekyLeveragingLarge2024,shestovFinetuningLargeLanguage2024,zhouLargeLanguageModel2024}, code generation \cite{nijkampCodeGenOpenLarge2022,friedInCoderGenerativeModel2022}, and optimization \cite{cumminsLargeLanguageModels2023}, heralding their immense potential for widespread application in software engineering \cite{zanLargeLanguageModels2023}. Although LLMs have not yet been directly applied to guide directed fuzzing, we have found that LLMs can offer more informative support for the orientation of directed fuzzing by understanding the structure and contextual semantics of code.

Although the internal workings of LLMs remain only partially understood, their deep code comprehension and analysis capabilities have been extensively validated in software engineering. The foundational attention mechanism in LLMs—originally inspired by human intuition models—first achieved remarkable success in neural machine translation. Conceptually, attention assigns a set of importance weights: when predicting a target element (e.g., a word in a sentence), the attention vector quantifies its relevance to every other element, and the resulting weighted sum yields a context-aware representation. Compared to natural language, program code places an even stronger emphasis on structural relationships among tokens, and many security vulnerabilities are tightly coupled to specific code patterns. By quantifying inter-token relevance and importance, attention scores naturally elevate potentially critical regions, enabling focused analysis of complex program logic.

DGF seeks to bias test inputs toward the code regions most likely to contain vulnerabilities, thereby accelerating fault discovery. However, existing structure-based distance metrics typically provide only a coarse “endpoint-distance” signal—much like knowing a maze’s exit direction without guidance on the optimal route. While useful, such metrics fail to capture the logical dependencies required to identify and traverse the most effective paths.

To address this gap, we introduce \textbf{Attention Distance}, a novel metric that fuses LLM-extracted semantic information with traditional fuzz-testing techniques to enhance DGF efficiency. Specifically, we employ a lightweight LLM, fine-tuned for directed fuzzing tasks, to compute an attention weight for each program statement and then aggregate these weights into a fine-grained distance guide. During the exploration phase, \textbf{Attention Distance} balances both structural context and semantic content to provide more precise path guidance. Our empirical evaluation across multiple real-world vulnerability scenarios demonstrates that, by simply replacing the existing distance metric—without modifying any other fuzzing components—we achieve substantial improvements in fault-finding effectiveness and coverage efficiency.

The primary contributions of this work are as follows:

\begin{itemize}

\item \textbf{Deep integration of LLM semantics with directed fuzzing}: To the best of our knowledge, this is the first effort to tightly couple LLM-extracted semantic insights with fuzz-testing guidance, and to validate its practical benefits.

\item \textbf{Introduction of the “Attention Distance” metric}: We propose a novel distance evaluation method that unifies code-structure and semantic information. By applying Attention Distance in directed fuzzing, we resolve the common issue of uniform distance values and markedly improve testing time efficiency.

\item \textbf{Modular implementation and open-source release}: We have developed a modular package that integrates seamlessly with existing directed fuzzing tools, leveraging the analytical power of LLMs to provide pinpoint path guidance. We will release our implementation and experimental datasets to support and stimulate further research in this area.
\end{itemize}

\section{Motivation}

In this section, we first examine the guidance distances used by existing directed fuzzing techniques and discuss their representational limitations. We then present a vulnerability example (see Figure \hyperref[fig:motivation]{3}) that highlights the key challenges and illustrates how the Attention Distance–guided method differs from conventional approaches.

\subsection{Distance Defects}

\begin{table*}[ht]
    \centering
    \caption{The content describes the distances between different Bug IDs and the proportion of these distances that occur most frequently relative to the total number of distances. Specifically, "Top 1" represents the proportion of occurrences for the most frequent distance, "Top 2" for the second most frequent distance, and "Top 3" for the third most frequent distance. "Top 1-3" denotes the combined proportion of occurrences for the three most frequent distances.}
    \resizebox{0.65\textwidth}{!}{ 

    \begin{tabular}{lcccccccccc}
    \toprule
    BUG ID & \makecell{Number of \\ Total Distances} & \makecell{Number of \\Unique Distances} & \makecell{Higest frequency\\ distance} & Top-1 & Top-2 & Top-3 & Top 1-3  \\
    \midrule
CVE-2016-9827 & 290 & 58 & 10.00 & 55.56\% & 12.76\% & 5.17\% & 73.45\%  \\
CVE-2017-9988 &23 & 18 & 20.00&13.04\%&8.70\%&8.70\%&30.43\%\\
CVE-2017-11728 & 297 & 118 & 30.00 & 19.95\% & 8.39\% & 5.67\% & 34.01\% \\
CVE-2018-7868 &645 &197 &20.00 &17.36\%&7.44\%&6.20\%&31.00\%\\
CVE-2018-8807 & 645 &192 &20.00 &12.09\%&9.46\%&7.13\%&28.68\%\\
CVE-2018-11095 & 582 &114 &30.00 &8.60\%&5.15\%&4.12\%&17.87\%\\
CVE-2019-9114 &680 & 209 &20.00 &16.03\%&9.41\%&6.91\%&32.35\%\\
binutils-2.31 & 131 & 37 &37.00 &18.32\%&6.87\%&6.11\%&31.30\%\\
binutils-2.29 & 239 & 37 & 20.00 & 27.62\% & 5.86\% & 4.18\% & 37.66\% \\
binutils-2.28 & 68 & 17 & 17.00 & 25.00\% & 14.71\% & 10.29\% & 50.00\% \\
LMS & 1105 & 557 & 13.33 & 24.34\% & 0.54\% & 0.27\% & 25.16\%\\
mjs-78 & 40&25 &20.00 &12.50\%&10.00\%&10.00\%&32.50\%\\
    \bottomrule
    \end{tabular}
    }
    \label{DistanceProblem}

\end{table*} 

\begin{figure}[!ht]
    \captionsetup{aboveskip=1pt, belowskip=1pt}

    \includegraphics[width=\linewidth]{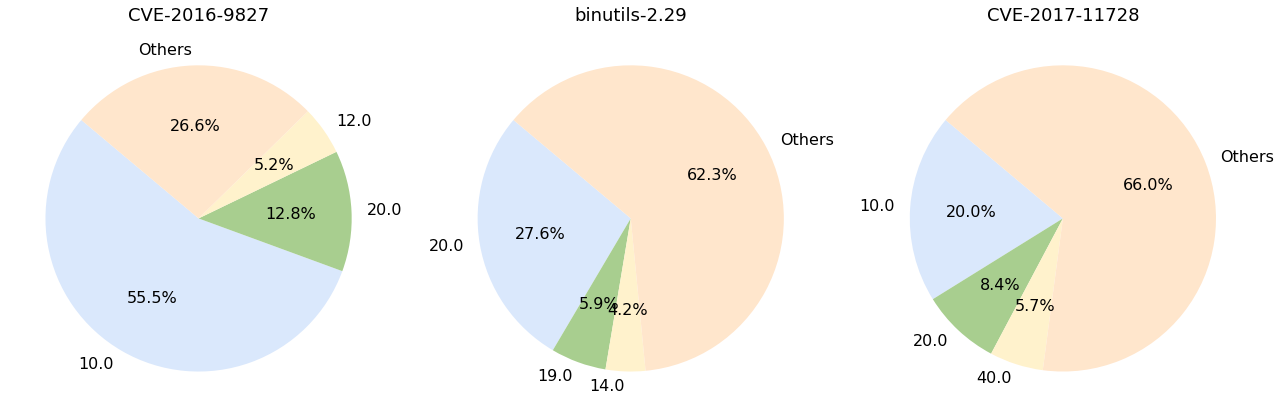}
    \caption{Pie charts depicting the distribution of distance values for bugs identified as CVE-2016-9827, binutils-2.29, and CVE-2017-11728. Each pie chart illustrates the proportion of the top three distances relative to the total distances observed for each bug.} 
    \label{fig:bingtu}
\end{figure}

In our replication tasks for real vulnerabilities using DGF, we identified critical deficiencies in the distance metrics. Our analysis of distance data, calculated by AFLGo and displayed in Table \ref{DistanceProblem}, reveals a prevalent issue of redundant distance values across numerous tasks. For instance, as depicted in Figure \ref{fig:bingtu}, in the analysis of the swftophp project pertaining to CVE-2016-9827, the most frequent distance value was 10, comprising 55.56\% of all recorded distances. Additionally, the top three most prevalent distances accounted for 73.54\% of the data, while the remaining 55 distance values constituted only 26.46\%. Similar patterns were observed in the binutils-2.29 project, where the most common distance was 30, accounting for 19.95\% of the distance data, and the top three distances summed up to 34.01\% of the total. These findings underscore a significant skew in the distribution of distance values across most projects, indicating a marked unevenness that could potentially impact the effectiveness of DGF in pinpointing vulnerabilities.

This indicates that in directed fuzzing, many basic‐block positions are uniformly prioritized. A primary cause of this uniformity is that conditional branch statements introduce parallel blocks within the code structure. However, current methodologies treat these parallel blocks as equivalent, overlooking how code logic impacts vulnerability triggering. During fuzzing, repeated distance values among such blocks can induce a temporary “blindness” in fuzzers, impeding their ability to effectively guide exploration toward target locations.

\subsection{Illustration of LLM guidance}

\begin{figure*}[!htbp]
    \captionsetup{aboveskip=1pt, belowskip=1pt}
    \centering
        \includegraphics[width=0.95\textwidth]{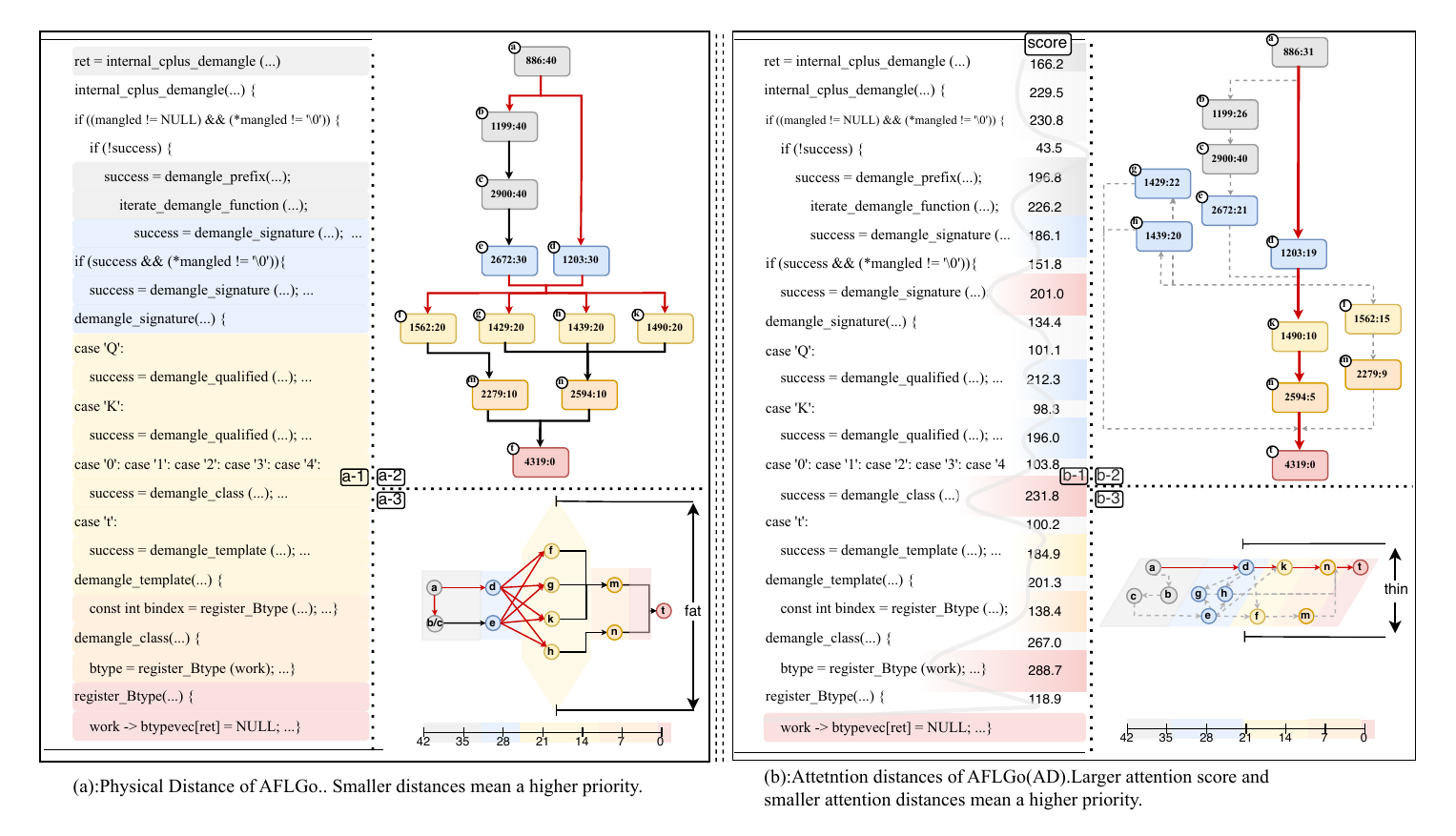}
        \caption{Motivational Example. We analyzed the CVE-2016-4487 example code using both AFLGo’s physical distances and our attention distances to compute metrics for critical code segments.Each node label follows the format “BlockID:Distance”, where the first number denotes the basic block identifier (corresponding to a specific code line), and the second indicates its computed distance to the target. From these metrics, we built the call-graph views shown in Figures (a-2) and (b-2), then abstracted the code-location relationships in Figures (a-3) and (b-3). Within the same code hierarchy, the attention method assigns noticeably higher scores to the critical segments, yielding more significant distance optimization. This advantage is clearly visible in the distance-relationship graphs, where the target segments stand out relative to other code units at the same level. }
        \label{fig:motivation}
    \end{figure*}

Although recent DGF tools like Beacon, WingRanger, and DAFL have made improvements in their guiding processes, their fundamental strategies still follow the design philosophy of AFLGo. Consequently, we categorize these tools as based on physical distances and use AFLGo as a representative example. To clarify this distinction, we present a simplified code snippet and call‐graph from the CVE-2016-4487 vulnerability to illustrate the differences between attention distance and the physical distance metric employed by AFLGo.

In the triggering path of CVE-2016-4487, multiple parallel conditional branches exist. Analysis begins at line 886 in the \textit{cplus-dem.c} file, which calls the \textit{internal\_cplus\_demangle} function. After a series of operations, the program enters the \textit{demangle\_signature} function, containing 23 parallel branches, but only a subset leads to the \textit{demangle\_class} function, triggering the vulnerability. For clarity, we have simplified the conditional branches and streamlined some of the code logic, as shown in Figure \ref{fig:motivation}.

Using the distance calculation method provided by AFLGo, we constructed a relationship graph between critical code blocks (see Figure \hyperref[fig:motivation]{3(a-2)}). This graph is based on basic block distances computed by AFLGo and attempts to categorize blocks with identical distances into the same hierarchical level. For example, in the file \textit{cplus\_dem.c}, lines 1203 and 2672 both invoke the \textit{demangle\_signature} function, which contains 23 branches. The figure demonstrates that the basic blocks at lines 1562, 1429, 1439, and 1490 all have a distance of 20. AFLGo's strategy prioritizes these blocks as high-priority seeds due to their proximity to the target, despite their varied contributions to triggering vulnerabilities. For instance, in \textit{cplus\_dem.c:1490}, the \textit{case '0'} triggers the \textit{register\_Btype} function via a call to \textit{demangle\_class}, eventually leading to vulnerability CVE-2016-4487. However, other branches such as \textit{case 'L'}, \textit{case 'C'}, \textit{case 'V'}, and \textit{case 'u'} do not trigger this vulnerability. If the fuzzer identifies two seeds simultaneously, where Seed \textbf{A} quickly enters the \textit{case 'L'} branch for execution and Seed \textbf{B} does not, AFLGo assigns a higher priority to Seed \textbf{A} even though it does not facilitate the triggering of the vulnerability. Additionally, as shown in Figure \hyperref[fig:motivation]{3(a-2)}, even though seeds triggering the \textit{e->f->m} path and those triggering the \textit{d->k->n} path are considered to have the same priority, the \textit{d->k->n} path is more exploitable for triggering the vulnerability.

To calculate the attention distance, we first compute attention scores for each line in the analyzed context code, as shown in Figure  \hyperref[fig:motivation]{3(b-1)} , where each line of code is scored within its context. Then, the guidance is directed based on these attention scores. We observe that the critical vulnerability triggering paths (1203, 1409, 2594) receive higher attention score evaluations among their code level peers. Thus, the distances of these critical codes can be assessed more precisely. Meanwhile, the system imposes penalties for irrelevant or erroneous function call paths (such as \textit{d->g}), ensuring that even if a seed penetrates deep into unrelated branches, it does not gain excessive priority weight. This ensures that DGF can explore towards the true targets without being misled by easily triggered erroneous branches.

Tools relying on physical distance strategies expose their limitations in handling complex programs because they only assess priority based on the call hierarchy, lacking deep analysis of code logic, despite some optimizations built on this basis. As shown in Figures \hyperref[fig:motivation]{3(a-3)} and Table \ref{DistanceProblem}, a plethora of identical distance indicators often lead to branch mazes and branch misguidance issues in many programs. These issues often result in the side effects of existing DGF, thereby reducing the effectiveness of the testing. Attention distance, by analyzing the logical relationships between codes and optimizing distance representations, not only resolves the flaws in distance assessment but also effectively handles issues caused by numerous branches with the same distances, as shown in Figure \hyperref[fig:motivation]{3(b-3)}, enhancing the overall performance of DGF. In the AFLGo framework, applying attention distance accelerates the reproduction of CVE-2016-4487 by a factor of 4.73 compared to physical distance guidance; detailed performance comparisons are provided in Section \ref{sectiomRQ2}.

\section{Design of the Attention Distance Metric}

\subsection{Overview}

\begin{figure*}[!ht]
  \captionsetup{aboveskip=1pt, belowskip=1pt}

  \centering
  \includegraphics[width=0.98\textwidth]{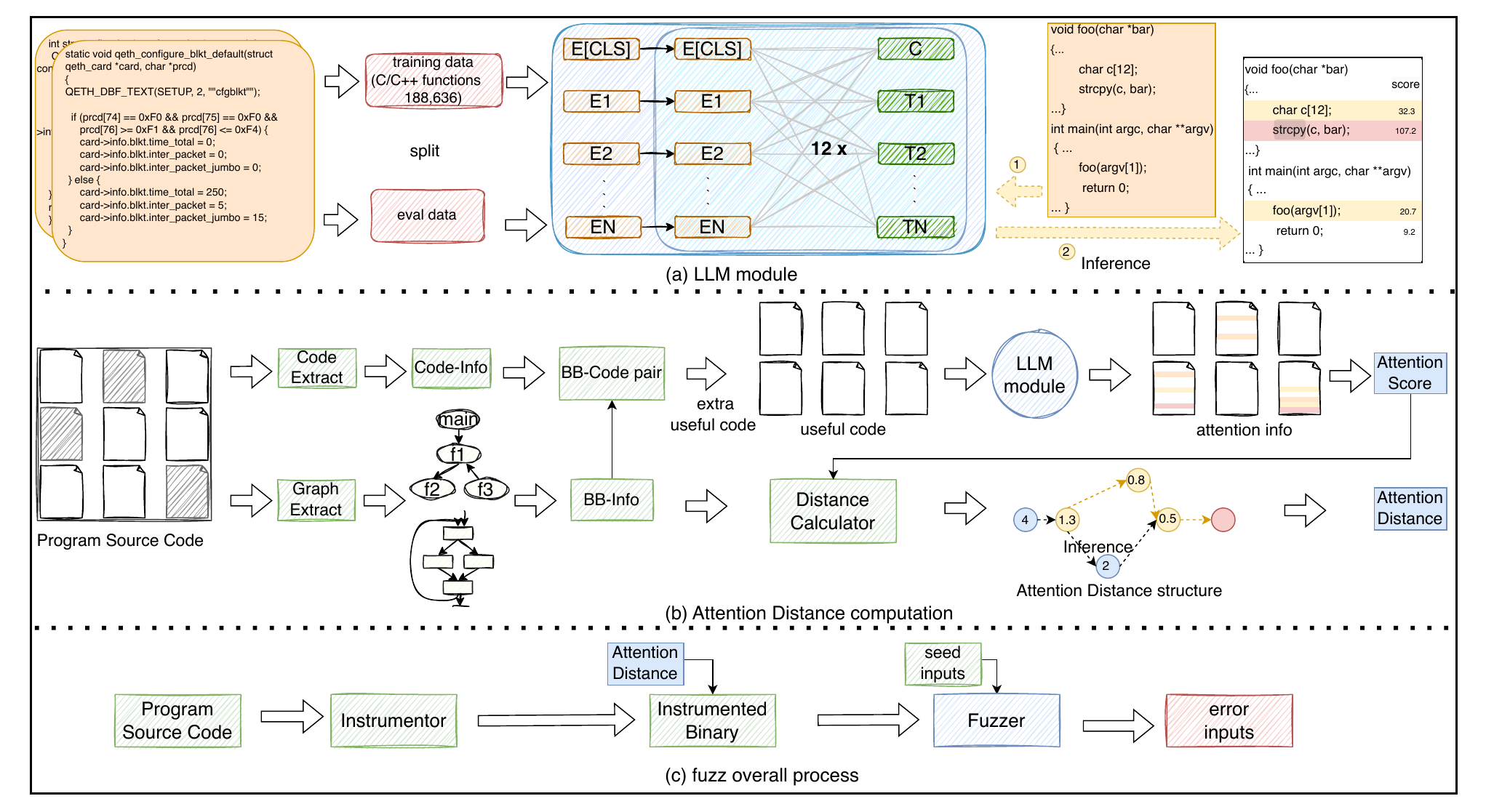}
  \caption{Architecture of the attention distance method. (a) Core LLM workflow: the model is trained on annotated vulnerability samples and, at inference, assigns attention scores to code segments. (b) Program processing: source code is split into basic-block pairs, which the LLM scores to derive attention distances. (c) Fuzzing workflow: the fuzzing engine leverages attention distances to guide directed fuzz testing.}
  \label{fig:framework}
\end{figure*}

Figure \hyperref[fig:framework]{4} illustrates the overall architecture of the attention-mechanism framework built upon AFLGo. Figure \hyperref[fig:framework]{4-a} provides a detailed overview of the LLM module’s training and inference pipeline: by learning from a vast collection of precisely annotated code samples, the module masters vulnerability recognition and associated pattern analysis, thereby enabling in-depth correlation analysis on code under test. Figure \hyperref[fig:framework]{4-b} reveals the operational principle of attention-augmented directed fuzz testing: the LLM assigns context-aware attention scores to each line of code, computes attention distances, and furnishes the fuzzing engine with path guidance. Figure \hyperref[fig:framework]{4-c} illustrates the full workflow of directed fuzz testing using attention distances.

\subsection{LLM module}
This section details the content displayed in Figure \hyperref[fig:framework]{4-a}. Conventional vulnerability identification models are generally trained on known vulnerability datasets with the goal of determining the presence of vulnerabilities through binary classification. However, this module does not solely focus on binary classification results but emphasizes identifying vulnerability associations at the code line level. The model conducts detailed reasoning and scoring of code blocks through its advanced attention mechanism and precisely describes the interconnections between codes by quantifying attention scores.

\subsubsection{Model Architecture Construction}

Our model architecture is based on CodeBERT \cite{devlinBERTPretrainingDeep} and incorporates the pre-training methodology of LineVul \cite{fuLineVulTransformerbasedLinelevel2022}. The choice of CodeBERT is primarily motivated by practical considerations and its balanced performance for code-related tasks. As a model pre-trained specifically on code, it enables us to leverage the existing fine-tuned LineVul checkpoint to directly obtain line-level vulnerability-related attention scores. This approach allows for efficient prototyping and computational feasibility, which is crucial for integration into fuzzing workflows where runtime performance is critical. Although more advanced or larger language models exist, they entail significantly higher computational costs, which would impede practical application in large-scale code repository analysis. Our research focus is on introducing semantic attention as a guidance mechanism for greybox fuzzing, rather than proposing a new model architecture.

The architecture itself is grounded in BERT, comprising 12 Transformer encoder layers, each equipped with bidirectional multi-head self-attention mechanisms and fully connected feed-forward networks. Through input embeddings, code sequences are converted into vector representations, with positional encoding added to preserve the sequence order. The core of the model is the multi-head self-attention mechanism, which enables the model to adaptively focus on different parts of the input sequence. By computing attention weights for each code element, the model forms targeted attention vectors that capture semantic relationships between code fragments.

\subsubsection{Attention Score}

In our framework, attention scores serve as a fundamental metric for quantifying the relationships among code elements. By employing multi-head attention, the model captures both global dependencies across the entire sequence and pinpoints the specific tokens most closely tied to security vulnerabilities. Processing multiple feature dimensions in parallel, this mechanism yields a rich, context-aware representation of code semantics.

At its core, the attention mechanism computes how each element in the input sequence attends to every other element. As formalized in Equation~\ref{formula-1}, we first take the dot product of the query and key vectors (Q and K) to obtain raw scores, then scale them by \(\sqrt{d_k}\) to mitigate gradient instability. A softmax transformation converts these scaled scores into a probability distribution of attention weights. Finally, these weights are applied to the value vectors (V), producing output vectors that encode the interactions and relative importance of all input elements \cite{vaswaniAttentionAllYou2017}.

\begin{equation}
  \begin{split}
  \text{Attention}(Q, K, V) = \text{softmax}\left(\frac{QK^T}{\sqrt{d_k}}\right)V
  \end{split}
  \label{formula-1}
\end{equation}

Multi‐head attention\cite{vaswaniAttentionAllYou2017} generalizes the self‐attention mechanism by simultaneously executing $h$ parallel attention heads, each operating in its own learned projection subspace to capture complementary inter‐token relationships. Formally, as shown in Equation \ref{formula-2}, the \texttt{MultiHead} operation concatenates the outputs of these $h$ heads and subjects the result to a final linear transformation:
\begin{equation}
  \text{MultiHead}(Q, K, V) = \text{Concat}(\text{head}_1, \dots, \text{head}_h)\;W^O
  \label{formula-2}
\end{equation}
where
\begin{equation}
  \text{head}_i = \text{Attention}\bigl(QW_i^Q,\; KW_i^K,\; VW_i^V\bigr)
\end{equation}
and each \(W_i^Q, W_i^K, W_i^V\) maps inputs into the \(i\)-th subspace. This design enables the model to attend to information from multiple representation spaces simultaneously, greatly enhancing its representational depth and flexibility.

In the Transformer-based code analysis model, we aggregate attention weights across all \(L\) encoder layers to compute a cumulative score for each code token. Let \(s_{ij}\) denote the attention weight of token \(i\) at layer \(j\). The line-level attention score is then computed by summing over tokens and layers:

\begin{equation}
  \text{LineScore} = \sum_{j=1}^{L}\;\sum_{i=1}^{N} s_{ij}
  \label{formula-3}
\end{equation}
where \(N\) is the number of tokens in that line. This aggregation captures both intra-line token interactions and their overall salience for vulnerability detection, thereby guiding the directed fuzzing engine toward the most semantically critical code regions~\cite{fuLineVulTransformerbasedLinelevel2022}.

Our framework employs attention-based scoring to quantify semantic relationships among code elements. The core insight stems from the behavior of vulnerability-tuned language models, which learn to assign higher attention weights to code tokens by analyzing both local syntax and broader contextual logic. We aggregate these token-level attention signals to produce line-level relevance scores, capturing the semantic importance of each code segment with respect to potential vulnerabilities. Although the aggregation process is methodologically straightforward, our principal contribution lies in repurposing the model’s contextualized attention distribution as a semantic relevance metric for fuzzing guidance. This approach injects rich, context-aware semantic understanding into vulnerability analysis, moving beyond purely syntactic or structural heuristics. The resulting scores offer fine-grained, interpretable insights that enhance automated security analysis. They provide empirical grounds for test case prioritization and vulnerability assessment, while also highlighting salient code tokens and structural patterns to focus fuzzing campaigns more effectively.

\subsubsection{Inference}

Our framework does not attempt to label code as “defective” directly; instead, it performs relational inference. Given a target program, we pass its code context into the LLM module, which conducts a detailed inference pass and assigns an attention weight to each token. These token-level weights are then aggregated into line-level attention scores, which subsequently drive the computation of attention distances for fuzzing guidance.

\subsection{Attention Distance computation}

This section elaborates on the computation of attention distance, as depicted in Figure~\hyperref[fig:framework]{4-b}. Once attention scores between code elements have been obtained, the system upgrades the traditional physical-distance metric by fusing it with semantic and structural insights. The resulting attention distance not only reflects static proximity within the program’s control-flow graph but also accounts for each element’s contextual relevance, yielding a richer measure of code relatedness. We begin by reviewing standard methods for calculating physical distance, then proceed to define and implement our attention-distance metric in detail.

\subsubsection{Physical Distance}

Distance measurement in directed fuzzing methods based on AFLGo, hereafter referred to simply as AFLGo, primarily relies on the distance between basic blocks within the Control Flow Graph (CFG), which is defined as physical distance. Compared to the complexity of computing the indirect control flow graph across the entire program, AFLGo concentrates on the precise calculation of target distances at the basic block level within a single process's CFG. This calculation is divided into two stages: initially determining the function-level distance followed by further refining the distance between basic blocks.

The function-level target distance assesses the distance from a function to all target functions in the call graph, while the function distance measures the distance between any two functions in the call graph. AFLGo defines the function distance \(d_f(n, n')\) as the number of edges on the shortest path between functions \(n\) and \(n'\) in the call graph CG, and the function-level target distance \(d_f(n, T_f)\) between function \(n\) and target function \(T_f\) as the harmonic mean of the function distance between \(n\) and any reachable target function \(tf \in T_f\):

\begin{equation}
  \begin{split}
  d_f(n,T_f) = 
  \begin{cases} 
  \text{undefined} & \text{if } R(n,T_f) = \emptyset \\
  \left( \sum_{t_f \in R(n,T_f)} \frac{1}{d_f(n,t_f)} \right)^{-1} & \text{otherwise}
  \end{cases}
  \end{split}
  \label{formula-4}
  \end{equation}

Basic block-level distance determines the separation between any two basic blocks within the CFG. The target distance at the basic block level, \( d_b(m, T_b) \), between a basic block \( m \) and the target basic block \( T_b \), is defined as follows:
\begin{equation}
\begin{split}
    db_{\text{phys}(m, T_b)} = \left\{
    \begin{array}{ll}
        0 & \text{if } m \in T_{b} \\
        c \cdot \min_{n \in N(m)}\left(d_{f}(n, T_{f})\right) & \text{if } m \in T \\
        \left[\sum_{t \in T}\left(d_{b}(m, t) + d_{b}(t, T_{b})\right)^{-1}\right]^{-1} & \text{otherwise}
    \end{array}
    \right.
\end{split}
\label{formula-5}
\end{equation}

\subsubsection{Attention Distance}

The calculation of attention distance is divided into two parts: at the function level and at the basic block level.

Due to the potential for excessively long inputs between function levels, the required focus on the code context may span a significant gap, potentially impacting the effectiveness of code relevancy analysis. Moreover, the instructional value of inter-function connectivity is less than that of line-by-line code connectivity. Therefore, we will not adjust our approach at the function level and will continue to use formula \ref{formula-4} to compute the distance between functions.

We compute the physical distances between basic blocks using formula \ref{formula-5}, then adjust these distances based on the attention scores of the blocks. The attention score for a basic block is defined as the sum of the scores of each line of code within the block, with individual line scores computed as specified in formula \ref{formula-3}. In this method, the original attention scores for each block \(m\), denoted as \(w_{\text{orig}}(m)\), are collected into the set \(W = \{w_{\text{orig}}(m) \mid m \in \text{Blocks}\}\). To prevent extreme scores—which may arise from the aggregation process—from disproportionately influencing the distance metric, we apply a soft capping mechanism. Specifically, the top 10\% of values in \(W\) are used as a threshold, and any score exceeding this cap is reduced to the threshold value. This strategy preserves the emphasis on highly attended regions while enhancing the robustness of the scoring distribution, as formalized below:

\begin{equation}
 \begin{split}
w_{\max} = \max_{m \in \text{Blocks}} \min\left(w_{\text{orig}}(m), W_{\left\lceil 0.1 \times |W| \right\rceil}\right)
\end{split}
\label{formula-6}
\end{equation}

The final computed normalized attention score \(w(m)\) is then calculated as follows:

\begin{equation}
 \begin{split}
w(m) =
\begin{cases}
0.5, & \text{if } w_{\min} = w_{\max} \\
\frac{\min\left(w_{\text{orig}}(m), W_{\left\lceil 0.1 \times |W| \right\rceil}\right) - w_{\min}}{w_{\max} - w_{\min}}, & \text{otherwise}
\end{cases}
\end{split}
\label{formula-7}
\end{equation}

 This normalization process yields processed attention scores that reflect the relative importance of each basic block. These normalized scores are then used to adjust the physical distances between blocks, culminating in the updated attention distance \( db_{\text{att}}(m, T_b) \):

\begin{equation}
  db_{att}(m, T_b) = db_{\text{phys}}(m, T_b) \times (S_a - w(m))  
  \label{formula-8}
 \end{equation}

where \(S_a\) is set at 1.5 to act as a constant parameter that scales the attention distance. This setting enhances the gradient of distance variations, thereby enabling a more intuitive and finely graded adjustment of distances. Shorter attention-based distances are applied to regions with high attention scores, emphasizing their importance, while areas with lower attention scores receive less focus, ensuring a balanced distribution of attention across the code.

\subsection{Implementation}

This implementation is built upon the AFLGo framework, with attention distance support introduced through targeted modifications to the LLM module and the fuzzing module.

For the LLM module, we fine-tuned the model to adapt it to the specific needs of fuzzing. We utilize the high-quality dataset constructed by Fan et al. \cite{fanCodeVulnerabilityDataset2020}, which meticulously annotates functions containing vulnerabilities and their specific line positions, significantly enhancing the model's precision in identifying vulnerability-related patterns in code and understanding the relationships between line-level code and other codes. The dataset includes 188,636 C/C++ functions from 348 open-source GitHub projects, containing 10,900 vulnerable functions (approximately 5.7\%), and 5,060,449 lines of code (with a similar proportion of vulnerable code lines, about 5.7\%), covering 91 CWE categories. The diversity and breadth of this dataset provide a rich sample base for model training, ensuring its effectiveness and accuracy in real-world applications.

To mitigate experimental errors caused by training data leakage, we conducted a thorough review of the code for the Time-to-Exposure experiments, using string searches to eliminate potential sources of leakage from the training dataset. To minimize random errors related to data splitting, we first employed an 80/20 split, where 80\% of the data was randomly assigned for training and 20\% for testing. Furthermore, to reduce evaluation bias from this partitioning, we utilized five-fold cross-validation. In this process, the dataset is randomly divided into five equal parts, with four parts used for training and the remaining one part reserved for validation. This procedure is repeated five times, ensuring each part serves as a test set once, which allows us to use the entire dataset for both training and evaluation, thus improving the robustness and generalizability of our results.

The core of fuzzing module module is the generation of attention distances, which involves three main steps. Given the high sensitivity of the LLM module to input content, we selectively filter code segments that have a practical impact on directed fuzzing, avoiding interference from comments and code processed during compilation. Initially, we modified LLVM PASS to capture program control flow graph information during the first compilation and extract valid code mappings corresponding to basic blocks. This step not only achieves effective mapping from basic blocks to code but also facilitates the extraction of valid code. Subsequently, we analyze these valid codes with the LLM module to obtain attention scores for each token and code line level. Finally, we use these scores to calculate the attention distances for various code positions. Modifications to LLVM PASS resulted in a reduction of about 200 lines of code, an addition of 1000 lines of C++ code, and over 1200 lines of Python code, used for invoking the LLM module for inference and calculating attention distances.For the entire fuzzing process, to better control variables and highlight the effects of attention distances, we adhere to the original design of AFLGo, and other parts of the design and implementation also follow AFLGo's settings.

Furthermore, to rigorously evaluate the general applicability of our attention-based scoring method, we have integrated it not only into the baseline AFLGo framework but also into two state-of-the-art directed fuzzers, DAFL and Windranger. A comprehensive comparative evaluation of these implementations will be presented in the subsequent experimental section.

\section{Evaluation}
In this section, we evaluate the Attention Distance method to verify its effectiveness and practical application value. Our analysis will focus on the following two core questions:

\begin{itemize}
  \item \textbf{RQ1:} Is attention distance more effective than traditional distance metrics at improving the efficiency of directed fuzz testing? 
  \item \textbf{RQ2} What underlying mechanisms enable attention distance to provide better directional guidance in directed fuzzing?
\end{itemize}

\textbf{Benchmark.} To ensure the reliability and accuracy of our comparative analysis, we utilized CVE data from the DAFL database as the test set. For each CVE, we independently established a test environment and replicated the vulnerability. Due to environmental constraints and other external factors, some vulnerabilities could not be successfully replicated. These data points were subsequently excluded from the final test set. Ultimately, we compiled a rigorously verified test set comprising 38 CVEs, the replication of which guarantees the quality and credibility of our experimental data.

\textbf{Experimental Setup.}All experiments were conducted in Docker on a 64-bit Ubuntu machine equipped with 12 CPU cores (Intel(R) Xeon(R) Gold 6126 CPU @ 2.60GHz), 125G of memory, and 1 GPU (24GB Nvidia GPU TU102 [TITAN RTX]) with CUDA 12.0. Each fuzzing session was run on a Docker container assigned with a single CPU core and 4GB of memory. 

\subsection{RQ1.Time-to-Exposure}
\label{sectiomRQ2}

\begin{table*}[!t]
  \centering
  \caption{Crash reproduction results of AFLGo(AD) and the baseline tools. The timeout is set to 12,000 seconds. MID represents the median time spent in fuzzing, and AVG represents the average time. Each statistic is the result of 50 repeated experiments. N.A.\ indicates that the tool could not produce valid output. * indicates that the fuzzer failed more than half of the trials; only successful runs are counted. “–” indicates unavailable results. Speedup is the ratio of baseline time to AD time for the same tool; higher is better.\textbf{Although DAFL(AD) and WindRanger(AD) outperform AFLGo(AD) in many scenarios, we use AFLGo(AD) as the representative configuration in our main performance analysis. This choice is made to isolate the effect of attention distance from tool-specific optimizations present in other fuzzers, ensuring a fair and focused evaluation of our proposed metric.}}
  \label{TTE}

  {%
  \setlength{\tabcolsep}{1.6pt}%
  \renewcommand{\arraystretch}{1.04}%

  \begin{adjustbox}{width=\textwidth}
    \begin{tabular}{l l
      c c   
      c c   
      c c   
      Y{2.4em} Y{2.2em}   
      c c   
      c c   
      Y{2.4em} Y{2.2em}   
      c c   
      c c   
      Y{2.4em} Y{2.2em}   
    }
    \toprule
    \textbf{Program} & \textbf{BUG ID}
    & \multicolumn{2}{c}{\textbf{AFL}}
    & \multicolumn{2}{c}{\textbf{AFLGo}}
    & \multicolumn{2}{c}{\textbf{AFLGo(AD)}}
    & \multicolumn{2}{c}{\textbf{\makecell{Speedup \\ \tiny(AFLGo(AD))}}}
    & \multicolumn{2}{c}{\textbf{DAFL}}
    & \multicolumn{2}{c}{\textbf{DAFL(AD)}}
    & \multicolumn{2}{c}{\textbf{\makecell{Speedup \\ \tiny(DAFL(AD))}}}
    & \multicolumn{2}{c}{\textbf{Windranger}}
    & \multicolumn{2}{c}{\textbf{Windranger(AD)}}
    & \multicolumn{2}{c}{\textbf{\makecell{Speedup \\ \tiny(WR(AD))}}}
    \\
    \cmidrule(lr){3-4}
    \cmidrule(lr){5-6}\cmidrule(lr){7-8}\cmidrule(lr){9-10}
    \cmidrule(lr){11-12}\cmidrule(lr){13-14}\cmidrule(lr){15-16}
    \cmidrule(lr){17-18}\cmidrule(lr){19-20}\cmidrule(lr){21-22}
    &
    & MID & AVG 
    & MID & AVG 
    & MID & AVG 
    & MID & AVG
    & MID & AVG 
    & MID & AVG 
    & MID & AVG
    & MID & AVG 
    & MID & AVG 
    & MID & AVG
    \\
    \midrule
        & 2016-9827
        & 73 & 98
        & 75 & 89
        & 66 & \textbf{67}
        & \cellcolor{LightGreen}1.14 & \cellcolor{LightGreen}1.33
        & 71 & 92
        & 25 & 24
        & \cellcolor{LightGreen}2.84 & \cellcolor{LightGreen}3.83
        & \textbf{61} & 70
        & 49 & 44
        & \cellcolor{LightGreen}1.24 & \cellcolor{LightGreen}1.59
    \\
        & 2016-9829
        & \textbf{264} & 633
        & 336 & 527
        & 481 & \textbf{462}
        & \cellcolor{LightRed}0.70 & \cellcolor{LightGreen}1.14
        & 446 & 515
        & 64 & 87
        & \cellcolor{LightGreen}6.97 & \cellcolor{LightGreen}5.92
        & 1458 & 1969
        & 805 & 1625
        & \cellcolor{LightGreen}1.81 & \cellcolor{LightGreen}1.21
    \\
        & 2016-9831
        & 371 & 514
        & 445 & 551
        & 296 & \textbf{358}
        & \cellcolor{LightGreen}1.50 & \cellcolor{LightGreen}1.54
        & \textbf{206} & 362
        & 111 & 101
        & \cellcolor{LightGreen}1.86 & \cellcolor{LightGreen}3.58
        & 1506 & 1222
        & N.A. & N.A.
        & -- & --
    \\
        & 2017-9988
        & 1817 & 2055
        & 1602 & 1908
        & \textbf{662} & \textbf{951}
        & \cellcolor{LightGreen}2.41 & \cellcolor{LightGreen}2.01
        & 1861 & 1879
        & N.A. & N.A.
        & -- & --
        & 1461 & 1721
        & N.A. & N.A.
        & -- & --
    \\
        & 2017-11728
        & 204 & 386
        & 185 & 361
        & N.A. & N.A.
        & -- & --
        & \textbf{98} & \textbf{160}
        & 243 & 225
        & \cellcolor{LightRed}0.40 & \cellcolor{LightRed}0.71
        & 790 & 958
        & 1013 & 998
        & \cellcolor{LightRed}0.78 & \cellcolor{LightRed}0.96
    \\
        & 2017-11729
        & 192 & 359
        & 180 & 389
        & 1849 & 1954
        & \cellcolor{LightRed}0.10 & \cellcolor{LightRed}0.20
        & 131 & \textbf{182}
        & 145 & 124
        & \cellcolor{LightRed}0.90 & \cellcolor{LightGreen}1.47
        & \textbf{101} & 220
        & 288 & 371
        & \cellcolor{LightRed}0.35 & \cellcolor{LightRed}0.59
    \\
        & 2017-7578
        & 357 & 570
        & 384 & 429
        & N.A. & N.A.
        & \cellcolor{LightGreen}1.49 & \cellcolor{LightGreen}1.74
        & 243 & 307
        & 129 & 137
        & \cellcolor{LightGreen}1.88 & \cellcolor{LightGreen}2.24
        & N.A. & N.A.
        & N.A. & N.A.
        & -- & --
    \\
        swftophp & 2018-7868
        & 5146 & 4983
        & 4610 & 5012
        & \textbf{1837} & \textbf{2029}
        & \cellcolor{LightGreen}2.51 & \cellcolor{LightGreen}2.47
        & *7568 & --
        & N.A. & N.A.
        & -- & --
        & *5696 & --
        & N.A. & N.A.
        & -- & --
    \\
        & 2018-8807
        & N.A. & N.A.
        & N.A. & N.A.
        & N.A. & N.A.
        & -- & --
        & N.A. & N.A.
        & N.A. & N.A.
        & -- & --
        & \textbf{*2888} & --
        & N.A. & N.A.
        & -- & --
    \\
        & 2018-8962
        & 3952 & \textbf{4158}
        & *3940 & --
        & N.A. & N.A.
        & -- & --
        & \textbf{*1782} & --
        & N.A. & N.A.
        & -- & --
        & *6951 & --
        & N.A. & N.A.
        & -- & --
    \\
        & 2018-11095
        & 3101 & 3037
        & 1692 & 2550
        & 1218 & 1309
        & \cellcolor{LightGreen}1.39 & \cellcolor{LightGreen}1.95
        & \textbf{441} & 1183
        & 383 & 376
        & \cellcolor{LightGreen}1.15 & \cellcolor{LightGreen}3.15
        & 734 & \textbf{871}
        & 501 & 552
        & \cellcolor{LightGreen}1.47 & \cellcolor{LightGreen}1.58
    \\
        & 2018-11225
        & N.A. & N.A.
        & N.A. & N.A.
        & N.A. & N.A.
        & -- & --
        & N.A. & N.A.
        & N.A. & N.A.
        & -- & --
        & N.A. & N.A.
        & N.A. & N.A.
        & -- & --
    \\
        & 2018-11226
        & N.A. & N.A.
        & N.A. & N.A.
        & *7953 & --
        & -- & --
        & \textbf{*6129} & --
        & 3263 & 3263
        & -- & --
        & N.A. & N.A.
        & N.A. & N.A.
        & -- & --
    \\
        & 2018-20427
        & *6749 & --
        & \textbf{*3321} & --
        & N.A. & N.A.
        & -- & --
        & N.A. & N.A.
        & 2969 & 2969
        & -- & --
        & *7906 & --
        & N.A. & N.A.
        & -- & --
    \\
        & 2019-12982
        & N.A. & N.A.
        & N.A. & N.A.
        & \textbf{*2050} & --
        & -- & --
        & \textbf{*2555} & --
        & 5636 & 5636
        & -- & --
        & N.A. & N.A.
        & N.A. & N.A.
        & -- & --
    \\
        & 2020-6628
        & N.A. & N.A.
        & N.A. & N.A.
        & N.A. & N.A.
        & -- & --
        & N.A. & N.A.
        & 6477 & 6477
        & -- & --
        & N.A. & N.A.
        & 3788 & 3834
        & -- & --
    \\
        \midrule
        & 2017-5969
        & 333 & 568
        & 344 & 388
        & \textbf{63} & \textbf{101}
        & \cellcolor{LightGreen}5.46 & \cellcolor{LightGreen}3.84
        & 217 & 441
        & 98 & 106
        & \cellcolor{LightGreen}2.21 & \cellcolor{LightGreen}4.16
        & 366 & 717
        & 336 & 336
        & \cellcolor{LightGreen}1.09 & \cellcolor{LightGreen}2.13
    \\
        xmllint & 2017-9047
        & N.A. & N.A.
        & N.A. & N.A.
        & \textbf{7274} & \textbf{7341}
        & -- & --
        & N.A. & N.A.
        & N.A. & N.A.
        & -- & --
        & N.A. & N.A.
        & N.A. & N.A.
        & -- & --
    \\
        & 2017-9048
        & N.A. & N.A.
        & N.A. & N.A.
        & N.A. & N.A.
        & -- & --
        & \textbf{3563} & \textbf{3966}
        & 1974 & 1974
        & \cellcolor{LightGreen}1.80 & \cellcolor{LightGreen}2.01
        & N.A. & N.A.
        & N.A. & N.A.
        & -- & --
    \\
        \midrule
        lrzip & 2017-8846
        & N.A. & N.A.
        & N.A. & N.A.
        & N.A. & N.A.
        & -- & --
        & N.A. & N.A.
        & N.A. & N.A.
        & -- & --
        & N.A. & N.A.
        & N.A. & N.A.
        & -- & --
    \\
        & 2018-11496
        & 25 & 27
        & 9 & 8
        & \textbf{6} & \textbf{8}
        & \cellcolor{LightGreen}1.50 & \cellcolor{LightGreen}1.00
        & 9 & 10
        & 8 & 5
        & \cellcolor{LightGreen}1.13 & \cellcolor{LightGreen}2.00
        & 10 & 10
        & 20 & 18
        & \cellcolor{LightRed}0.50 & \cellcolor{LightRed}0.56
    \\
        \midrule
        & 2016-4487
        & 651 & 697
        & 645 & 776
        & \textbf{149} & \textbf{164}
        & \cellcolor{LightGreen}4.33 & \cellcolor{LightGreen}4.73
        & 271 & 293
        & 257 & 211
        & \cellcolor{LightGreen}1.05 & \cellcolor{LightGreen}1.39
        & 572 & 548
        & 258 & 255
        & \cellcolor{LightGreen}2.22 & \cellcolor{LightGreen}2.15
    \\
        & 2016-4489
        & 1262 & 1329
        & 773 & 981
        & \textbf{282} & \textbf{374}
        & \cellcolor{LightGreen}2.74 & \cellcolor{LightGreen}2.62
        & 852 & 1129
        & 512 & 798
        & \cellcolor{LightGreen}1.66 & \cellcolor{LightGreen}1.41
        & 428 & 448
        & 294 & 327
        & \cellcolor{LightGreen}1.46 & \cellcolor{LightGreen}1.37
    \\
        cxxfilt & 2016-4490
        & 387 & 392
        & 343 & 437
        & 84 & 93
        & \cellcolor{LightGreen}4.08 & \cellcolor{LightGreen}4.70
        & \textbf{72} & \textbf{83}
        & 48 & 48
        & \cellcolor{LightGreen}1.50 & \cellcolor{LightGreen}1.73
        & 206 & 226
        & 142 & 161
        & \cellcolor{LightGreen}1.45 & \cellcolor{LightGreen}1.40
    \\
        & 2016-4491
        & *5833 & *5795
        & *5132 & *5127
        & *1010 & \textbf{*953}
        & \cellcolor{LightGreen}5.08 & \cellcolor{LightGreen}5.38
        & \textbf{*925} & *996
        & N.A. & N.A.
        & -- & --
        & *5642 & *5518
        & N.A. & N.A.
        & -- & --
    \\
        & 2016-4492
        & 676 & 693
        & 899 & 967
        & 295 & 371
        & \cellcolor{LightGreen}3.05 & \cellcolor{LightGreen}2.61
        & 322 & 378
        & 254 & 245
        & \cellcolor{LightGreen}1.27 & \cellcolor{LightGreen}1.54
        & 963 & 964
        & 933 & 937
        & \cellcolor{LightGreen}1.03 & \cellcolor{LightGreen}1.03
    \\
        & 2016-6131
        & 1637 & 1833
        & 1994 & 2292
        & 311 & 420
        & \cellcolor{LightGreen}6.41 & \cellcolor{LightGreen}5.46
        & 513 & 591
        & N.A. & N.A.
        & -- & --
        & 5161 & 4696
        & N.A. & N.A.
        & -- & --
    \\
        \midrule
        & 2017-8393
        & 1927 & 2419
        & 993 & 1152
        & N.A. & N.A.
        & -- & --
        & \textbf{635} & \textbf{1004}
        & 172 & 251
        & \cellcolor{LightGreen}3.69 & \cellcolor{LightGreen}4.00
        & 1492 & 1407
        & 426 & 410
        & \cellcolor{LightGreen}3.50 & \cellcolor{LightGreen}3.43
    \\
        objcopy & 2017-8394
        & 845 & 937
        & 684 & 775
        & 196 & 174
        & \cellcolor{LightGreen}3.49 & \cellcolor{LightGreen}4.45
        & 210 & 314
        & 271 & 245
        & \cellcolor{LightRed}0.77 & \cellcolor{LightGreen}1.28
        & 612 & 900
        & 212 & 221
        & \cellcolor{LightGreen}2.89 & \cellcolor{LightGreen}4.07
    \\
        & 2017-8395
        & 118 & 120
        & 113 & 118
        & 18 & 19
        & \cellcolor{LightGreen}6.28 & \cellcolor{LightGreen}6.21
        & \textbf{15} & \textbf{16}
        & 11 & 10
        & \cellcolor{LightGreen}1.36 & \cellcolor{LightGreen}1.60
        & 113 & 119.5
        & 166 & 164
        & \cellcolor{LightRed}0.68 & \cellcolor{LightRed}0.73
    \\
        \midrule
        & 2017-8392
        & 326 & 335
        & 354 & 374
        & N.A. & N.A.
        & -- & --
        & \textbf{35} & \textbf{75}
        & 21 & 19
        & \cellcolor{LightGreen}1.67 & \cellcolor{LightGreen}3.95
        & N.A. & N.A.
        & N.A. & N.A.
        & -- & --
    \\
        & 2017-8396
        & N.A. & N.A.
        & N.A. & N.A.
        & N.A. & N.A.
        & -- & --
        & N.A. & N.A.
        & N.A. & N.A.
        & -- & --
        & N.A. & N.A.
        & N.A. & N.A.
        & -- & --
    \\
        objdump & 2017-8397
        & \textbf{*7041} & --
        & N.A. & N.A.
        & N.A. & N.A.
        & -- & --
        & *9610 & --
        & 4834 & 4834
        & -- & --
        & N.A. & N.A.
        & N.A. & N.A.
        & -- & --
    \\
        & 2017-8398
        & 1809 & 2731
        & 2456 & 3538
        & 347 & 866
        & \cellcolor{LightGreen}7.08 & \cellcolor{LightGreen}4.09
        & 499 & \textbf{469}
        & 432 & 339
        & \cellcolor{LightGreen}1.16 & \cellcolor{LightGreen}1.38
        & N.A. & N.A.
        & 218 & 315
        & -- & --
    \\
        & 2018-17360
        & N.A. & N.A.
        & N.A. & N.A.
        & N.A. & N.A.
        & -- & --
        & N.A. & N.A.
        & N.A. & N.A.
        & -- & --
        & N.A. & N.A.
        & N.A. & N.A.
        & -- & --
    \\
        \midrule
        nm & 2017-14940
        & N.A. & N.A.
        & 993 & 1152
        & N.A. & N.A.
        & -- & --
        & \textbf{635} & \textbf{1004}
        & N.A. & N.A.
        & -- & --
        & 1492 & 1407
        & N.A. & N.A.
        & -- & --
    \\
        \midrule
        readelf & 2017-16828
        & 171 & 282
        & 116 & 227
        & 45 & 45
        & \cellcolor{LightRed}0.52 & \cellcolor{LightRed}0.97
        & 271 & 553
        & 31 & 32
        & \cellcolor{LightGreen}8.74 & \cellcolor{LightGreen}17.28
        & \textbf{36} & \textbf{55}
        & 45 & 45
        & \cellcolor{LightRed}0.80 & \cellcolor{LightGreen}1.22
    \\
        \midrule
        strip & 2017-7303
        & 123 & 133
        & 128 & 144
        & \textbf{17} & \textbf{18}
        & \cellcolor{LightGreen}7.53 & \cellcolor{LightGreen}8.00
        & 288 & 412
        & 91 & 90
        & \cellcolor{LightGreen}3.16 & \cellcolor{LightGreen}4.58
        & 1361 & 1518
        & 308 & 413
        & \cellcolor{LightGreen}4.42 & \cellcolor{LightGreen}3.68
    \\
    \bottomrule
    \end{tabular}
  \end{adjustbox}
  }%

\end{table*}

For a fair comparison of attention‐based distances, we built AFLGo-AD on top of AFLGo and used attention distances to guide directed fuzzing without modifying any other fuzzing components (e.g., seed selection or block optimizations). In our crash reproduction experiments to evaluate time‐to‐error (TTE) performance, we conducted 50 independent trials for each CVE. If the fuzzer failed to reproduce the target crash within 12,000 seconds in more than half of those trials, we marked the result as not applicable (N.A.). For CVEs that were successfully reproduced within the time limit, we report both the mean and median reproduction times; for cases with only partial reproducibility, we report the median time to avoid skewing by outliers.

As shown in Table \ref{TTE}, across 38 real CVEs, simply replacing the guidance distance with attention distance yielded significant gains in 19 of the 21 vulnerabilities that both AFLGo and AFLGo-AD could consistently reproduce: median reproduction time was \textbf{3.58×} faster and average reproduction time \textbf{3.43×} faster. Excluding the five CVEs that no method could trigger within 12,000 s, AFLGo-AD achieved the best median time on 14 vulnerabilities and the best average time on 16 vulnerabilities, outperforming all other mainstream directed fuzzers. Moreover, in median reproduction time AFLGo-AD was \textbf{2.07×} faster than DAFL and \textbf{7.28×} faster than Windranger, and in average reproduction time it was \textbf{2.89×} faster than DAFL and \textbf{7.13×} faster than Windranger.

Taking CVE-2017-7303 as a case study, we analyze how attention distances guide directed fuzzing more effectively. The vulnerability resides in the function \textbf{section\_match} (elf.c:1258), which is invoked by \textbf{find\_link} (elf.c:1321) and in turn called by \\ \textbf{copy\_special\_section\_field}. These functions form the core execution path, with multiple conditional branches that can trigger the bug. By integrating code-context analysis, attention distances refine focus on these critical functions: for example, the distance for \textbf{find\_link} is reduced from AFLGo’s default of 10 to a targeted 5. Moreover, they streamline the representation of all upstream branches related to \textbf{section\_match}, sharpening the fuzzer’s attention on that key segment. These strategic adjustments enable the engine to concentrate more quickly and efficiently on the branches most likely to expose the flaw, delivering substantial efficiency gains. In practical tests on CVE-2017-7303, simply replacing physical distances with attention distances made median and mean time-to-error 7.53× and 8.00× faster than AFLGo, 16.94× and 22.89× faster than DAFL, and 80.06× and 84.33× faster than Windranger. Similarly, in the CVE-2016-4487 example from our motivation section, attention-based guidance alone yielded 4.33× and 4.73× speedups in median and average reproduction times.

Based on our experimental analysis, we note that for several CVEs—including CVE-2017-11728, CVE-2017-9048, CVE-2017-8392, CVE-2017-14940, and CVE-2017-16828—the attention-guided approach did not reproduce crashes within the 12,000-second time limit, while some baseline tools succeeded. We attribute these outliers primarily to what we term “spurious attention peaks,” where the model assigns misleadingly high weights to code lines that are semantically salient but not relevant to the actual vulnerability—such as generic error handlers, common parsing routines, or keywords that distract from the true defect location. In other cases, especially in programs with large basic blocks or numerous functions, the attention distribution becomes overly flat, failing to sufficiently highlight vulnerability-critical paths. These observations reveal inherent challenges in relying solely on LLM-derived attention, particularly when the model misinterprets code semantics or lacks sharp focus. They also point to potential refinements, such as incorporating outlier detection or attention calibration, to reduce misguidance and strengthen the practical utility of semantic-aware fuzzing.

Furthermore, we integrate Attention Distance into two representative directed fuzzing tools—DAFL and Windranger—to evaluate its generalizability across different system architectures. It is worth noting that both tools already incorporate some form of semantic or logical distance metric, whose underlying modeling mechanisms could theoretically overlap or interfere with our attention-based measure. Nevertheless, experimental results demonstrate that Attention Distance still consistently improves overall performance in these already-optimized frameworks.After integrating Attention Distance, DAFL shows an average speedup of 2.25× in MID and 3.30× in AVG for crash reproduction time. Similarly, Windranger achieves an average speedup of 1.61× in MID and 1.73× in AVG. These results indicate that Attention Distance maintains a performance advantage on most benchmarks, even in the presence of sophisticated pre-existing semantic distance mechanisms. This not only confirms its portability as a general semantic measure, but also suggests a promising direction for enhancing directed fuzzing: leveraging the intrinsic attention distribution from pre-trained models to refine path priority estimation.

\vspace{4pt} 

\noindent\fcolorbox{black}{gray!30}{
  \parbox{\dimexpr\columnwidth-2\fboxsep-2\fboxrule\relax}{

AFLGo-AD shows improved performance in both average and median time metrics.  Compared to AFLGo, DAFL, and Windranger, its times for generating crashes are 3.13 and 2.07 times faster, 7.28 and 2.89 times faster, and 7.13 and 2.89 times faster, respectively, for median and average times.These significant improvements, achieved even when integrating Attention Distance into existing high-performing fuzzers, further validate its generalizability and practical value across diverse fuzzing architectures. 
  }
}

\subsection{RQ2.Guidance-Mechanism}

In this section, we evaluate the guiding effect of the attention mechanism.

\begin{figure*}[!ht] \centering
    \captionsetup{aboveskip=0pt, belowskip=0pt}  
    \includegraphics[width=\textwidth]{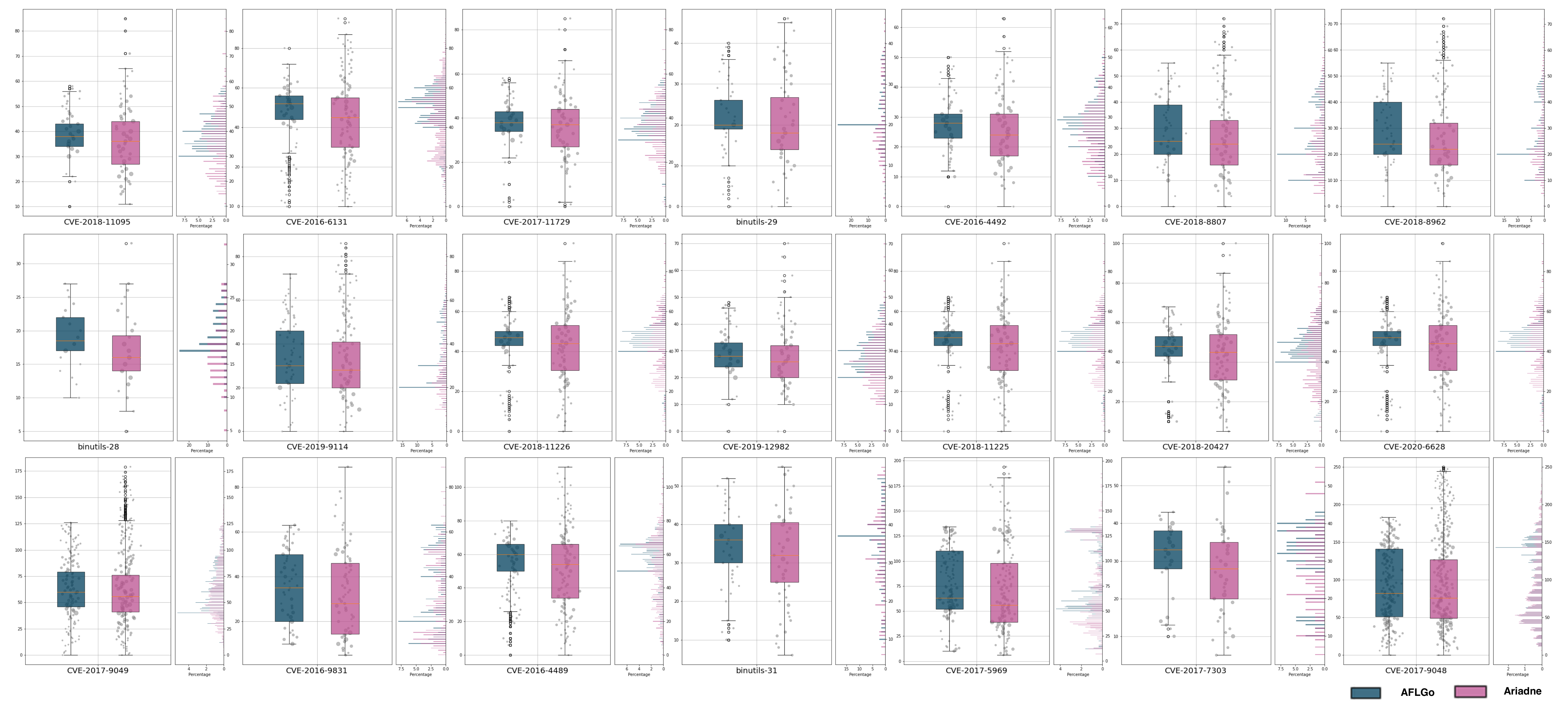}
    \caption{Basic‐block‐level distance distributions for physical distance and attention distance across different projects. In each subplot, the y-axis shows distance values via box plots with overlaid scatter points for the two metrics, while the bar chart on the right indicates the percentage frequency of each distance value. Only a portion of the binary is shown here; full results and supporting data are provided in the supplementary materials.}
    \label{fig:distance_improve}
\end{figure*}

\subsubsection{Distance improvement}

We conducted a systematic comparison between attention-based distances and traditional physical distances across 38 representative real-world vulnerabilities in binary programs. Figure \ref{fig:distance_improve} presents the results: the box plots summarize each metric’s key distribution statistics, the overlaid scatter plots visualize the frequency of individual distance values (with point density and size indicating occurrence frequency), and the adjacent histogram further highlights differences in distribution shape.

The experiments show that attention-based distances substantially alleviate the clustering seen with physical distances, producing a more layered and informative distribution. For example, in binutils-2.29, physical distances are highly concentrated—28\% of values fixed at 20 and an interquartile range tightly bound between 19 and 26. After applying attention distances, the IQR expands from 14 to 27, and the overall distribution becomes more dispersed and discriminative.This improvement arises because traditional physical distances rely solely on structural information—such as node positions in the control-flow graph—and cannot capture deep semantic relationships. Complex conditional-branch structures often yield many basic blocks that are structurally adjacent but semantically distinct. In contrast, attention distances extract context-aware attention scores between code segments, enabling a finer-grained, semantically informed, multivariate guidance strategy.

Overall, attention distances not only balance the distribution more evenly but also introduce a clearer gradient, reducing redundant stacking of data points at identical physical distances and significantly enhancing the discriminative power of directed guidance as well as fuzzing efficiency.

\subsubsection{Key Code Recognition}

To evaluate the ability of attention scores to identify and prioritize vulnerable code, we conducted systematic experiments at three granularity levels: function, basic-block, and code-line. Results are reported using the following metrics:
\begin{itemize}
  \item \textbf{Recall}: the proportion of true positives correctly identified among all actual positives.
  \item \textbf{Precision}: the proportion of true positives among all predicted positives.
  \item \textbf{F1 Score}: the harmonic mean of precision and recall.
\end{itemize}

\paragraph{Function-Level Recognition}

\begin{table}[!htb]
    \captionsetup{aboveskip=1pt, belowskip=1pt} 
    \centering
    \caption{Function-Level Discrimination Results}
    \resizebox{0.95\columnwidth}{!}{ 
    \begin{tabular}{cccccc}
    \toprule 
    Fold & Num Examples &  F1 &  Precision &  Recall & Total Inspected Lines \\
    \midrule 
    0 & 37728 & 0.9076 & 0.9618 & 0.8592 & 130015 \\
    1 & 37727 & 0.9094 & 0.9610 & 0.8631 & 129093 \\
    2 & 37727 & 0.9152 & 0.9688 & 0.8672 & 136564 \\
    3 & 37727 & 0.8826 & 0.9115 & 0.8556 & 136962 \\
    4 & 37727 & 0.8959 & 0.9579 & 0.8415 & 126337 \\
    \bottomrule 
    \end{tabular}
    }
    \label{table:linevul_results}
\end{table}

We first measured the model’s ability to flag vulnerable functions. To mitigate any bias from training–set distribution, we employed five-fold cross-validation. Table~\ref{table:linevul_results} shows the number of samples and performance metrics for each fold. Across all folds, F1 scores ranged from 0.8826 to 0.9152, precision from 0.9115 to 0.9688, and recall from 0.8415 to 0.8672. These stable results demonstrate that attention scores reliably pinpoint functions containing real vulnerabilities.

\paragraph{Block and Line-Level Recognition}

\begin{figure}[!ht]
    \captionsetup{aboveskip=0pt, belowskip=0pt} 
    \centering
    \includegraphics[width=\columnwidth]{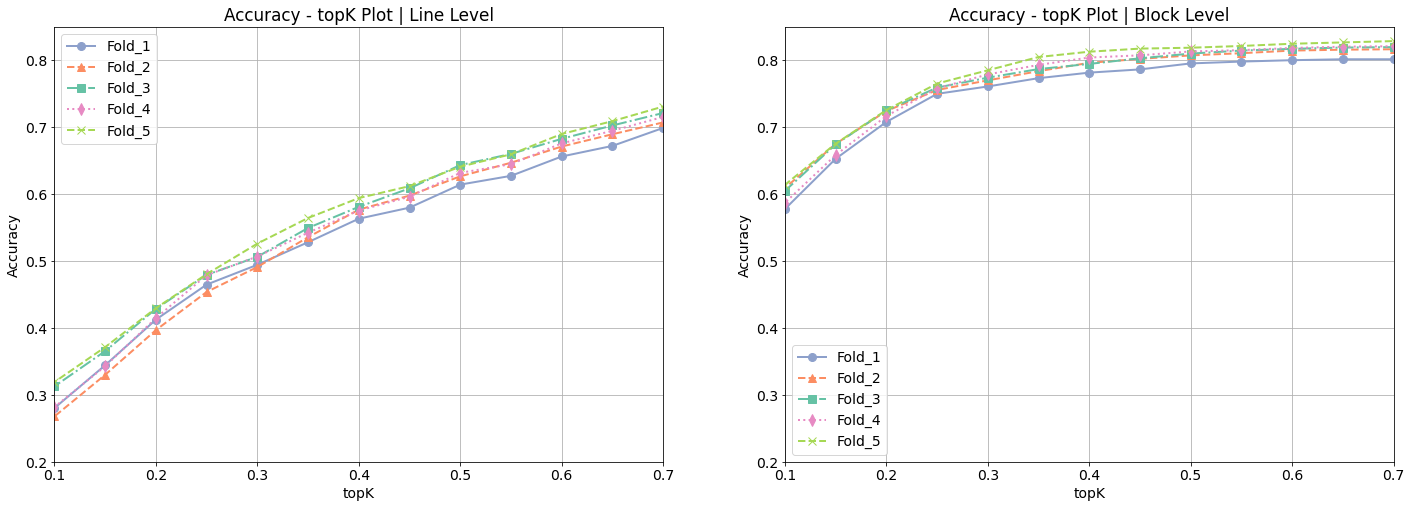}
    \caption{The accuracy in identifying basic blocks and lines. Critical basic blocks and key lines are determined solely by aggregating attention scores.}
    \label{fig:figure6}
\end{figure}
Next, we performed a fine-grained analysis by ranking basic blocks and individual lines according to their attention scores. As Figure~\ref{fig:figure6} illustrates, high-scoring regions consistently coincide with actual defect sites. At the block level, a Top-K threshold of 0.2 yields over 70\% accuracy, and raising the threshold to 0.4 approaches 80\%. At the line level, a 0.4 threshold achieves nearly 60\% accuracy. These findings confirm that aggregating token-level attention effectively highlights precise vulnerability locations.

\vspace{4pt} 
\noindent\fcolorbox{black}{gray!30}{
  \parbox{\dimexpr\columnwidth-2\fboxsep-2\fboxrule\relax}{
The attention-guided mechanism enhances directed fuzzing by jointly improving distance modeling and vulnerability localization. Attention-based distances yield more informative and balanced distributions, reducing redundancy and refining path guidance. Meanwhile, high attention scores effectively highlight vulnerable functions, blocks, and lines, demonstrating strong alignment between semantic relevance and defect likelihood. This unified mechanism enables more precise and efficient fuzzing through both structural and semantic cues.
  }
}

\section{Related Work}

Grey-box fuzzing has been proven as an effective technique for discovering software vulnerabilities in the real world \cite{choiSMARTIANEnhancingSmart2021,kimDriveFuzzDiscoveringAutonomous2022,liuStateSelectionAlgorithms2022,hanCodeAlchemistSemanticsawareCode2019,fanCodeVulnerabilityDataset2020,luoWestworldFuzzingassistedRemote2021} , with Coverage-Guided Fuzzing (CGF) particularly emphasizing the importance of guiding the fuzzing process to cover more code. AFL \cite{zalewski2021afl} stands as a classic example within this strategy, utilizing coverage extent as feedback to guide the evolution of seed inputs. Steelix \cite{liSteelixProgramstateBased2017} optimizes CGF path exploration through lightweight static analysis and binary instrumentation, while Angora \cite{chenAngoraEfficientFuzzing2018} adopts gradient descent-based input search techniques as an alternative to symbolic execution for solving constraints. FairFuzz \cite{lemieuxFairFuzzTargetedMutation2018} identifies and masks bytes in the generated inputs that affect the coverage of rare branches. AFLFast \cite{bohmeCoveragebasedGreyboxFuzzing2019} enhances path testing efficiency by concentrating fuzzing efforts on less frequently traversed paths. Vuzzer \cite{chenHawkeyeDesiredDirected2018a} prioritizes paths that are more likely to reveal vulnerabilities by assigning weights to specific basic blocks.

While CGF aims for broad code coverage, DGF focuses on identifying and detecting specific vulnerabilities in software by finely tuning test case generation strategies with distance feedback \cite{canakciTargetFuzzUsingDARTs2022,shahMC2RigorousEfficient2022}. Early implementations like AFLGo \cite{bohmeDirectedGreyboxFuzzing2017} have demonstrated the effectiveness of using this strategy to guide test cases towards predefined target locations, significantly improving testing specificity and efficiency. Subsequent works, such as Hawkeye \cite{chenHawkeyeDesiredDirected2018a} and WindRanger\cite{duWindRangerDirectedGreybox2022}, introduced more detailed information feedback and distance calculation mechanisms to further enhance the performance of directed fuzzing. Meanwhile, methods like Beacon \cite{huangBEACONDirectedGreybox2022}, SieveFuzz \cite{srivastavaOneFuzzDoesn2022} , and DAFL \cite{kimDAFLDirectedGreybox2023}, which employ program pruning and selective coverage feedback, have further increased the efficiency and effectiveness of DGF. SELECTFUZZ \cite{luoSelectFuzzEfficientDirected2023} accelerates directed exploration by selectively detecting and exploring code blocks related to the target.ParmeSan\cite{osterlundParmeSanSanitizerguidedGreybox2020} considers sanitizer guided bug coverage inaddition to the distance feedback. Additionally, FuzzGuard \cite{zongFuzzGuardFilteringOut2020} filters input seeds that cannot reach target sites using deep learning technology, further improving testing efficiency and outcomes.Directed White-box Fuzzing, or DirectedSymbolic Execution\cite{maDirectedSymbolicExecution2011,marinescuKATCHHighcoverageTesting2013,yangDirectedIncrementalSymbolic2014,christakisGuidingDynamicSymbolic2016,cadarKLEEUnassistedAutomatic2008} is a predecessor of DGF.

\section{Discussion and future work}

Our work focuses on enhancing the distance metric in directed fuzzing by incorporating semantic understanding derived from large language models. Technically, our approach represents an incremental yet novel contribution: rather than developing a new program analysis framework or complex scheduling algorithm, we are the first to integrate the semantic knowledge encoded in pretrained models into fuzzing guidance. We recognize that attention distance alone may not surpass all aspects of more sophisticated frameworks in every scenario. Accordingly, we position it not as a wholesale replacement, but as a complementary technique that can be integrated with existing methods. The strength of our approach lies in its emphasis on semantic relevance, which proves particularly useful when targeting vulnerabilities involving complex semantic conditions or cross-function logic—cases where purely syntactic or structural metrics often fall short. For example, when triggering a vulnerability depends on deep data dependencies or semantic constraints, static or structural distances may be inadequate, while the attention mechanism in a pretrained model can capture subtle inter-code relationships, providing our method with a systematic advantage in such situations.

While attention mechanisms excel at capturing relationships among code tokens and highlighting salient elements, they can be misled by atypical naming conventions or overly long lines. For instance, a function named `StackOverflow` might erroneously receive high attention scores, even when it is not related to a defect, simply because its name resembles a well-known vulnerability source. Such spurious attention peaks can distort the model’s assessment of code relevance. To mitigate this, future work should investigate refined token-scoring strategies—perhaps by normalizing or contextualizing token representations—to ensure that attention truly reflects semantic importance. Moreover, guidance in directed fuzzing need not be confined to distance metrics alone: enhancements in seed selection, branch pruning, and mutation strategies all present opportunities for further gains. Although this study focuses on a single, novel distance-based insight, synergistic optimizations across multiple fuzzing components could yield even greater performance improvements.

\section{Conclusion}
In this paper, we introduce a novel, attention-based distance metric—\textbf{Attention Distance}—to overcome the paucity of informative guidance in directed grey-box fuzzing. By leveraging a lightweight, task-tuned LLM to compute fine-grained attention weights for each program statement, we transform those weights into a contextual distance guide that supplants traditional physical-distance measures. Our experiments demonstrate that a drop-in replacement of the original metric in AFLGo yields dramatic improvements in vulnerability reproduction efficiency, surpassing state-of-the-art directed fuzzers such as AFL, AFLGo, WindRanger, and DAFL. Beyond boosting fuzzing performance, Attention Distance pioneers a new paradigm for fusing LLM-derived semantic insights with classic fuzz-testing workflows, opening fresh avenues for future research and practical application.

\begin{acks}
To Robert, for the bagels and explaining CMYK and color spaces.
\end{acks}

\bibliographystyle{ACM-Reference-Format}
\bibliography{acmartbib}

@inproceedings{bohmeDirectedGreyboxFuzzing2017,
  title={Directed greybox fuzzing},
  author={B{\"o}hme, Marcel and Pham, Van-Thuan and Nguyen, Manh-Dung and Roychoudhury, Abhik},
  booktitle={Proceedings of the 2017 ACM SIGSAC conference on computer and communications security},
  pages={2329--2344},
  year={2017}
}

@inproceedings{chenHawkeyeDesiredDirected2018a,
  title={Hawkeye: Towards a desired directed grey-box fuzzer},
  author={Chen, Hongxu and Xue, Yinxing and Li, Yuekang and Chen, Bihuan and Xie, Xiaofei and Wu, Xiuheng and Liu, Yang},
  booktitle={Proceedings of the 2018 ACM SIGSAC conference on computer and communications security},
  pages={2095--2108},
  year={2018}
}

@inproceedings{xuanCrashReproductionTest2015,
  title={Crash reproduction via test case mutation: Let existing test cases help},
  author={Xuan, Jifeng and Xie, Xiaoyuan and Monperrus, Martin},
  booktitle={Proceedings of the 2015 10th joint meeting on foundations of software engineering},
  pages={910--913},
  year={2015}
}

@inproceedings{shahMC2RigorousEfficient2022,
  title={Mc2: Rigorous and efficient directed greybox fuzzing},
  author={Shah, Abhishek and She, Dongdong and Sadhu, Samanway and Singal, Krish and Coffman, Peter and Jana, Suman},
  booktitle={Proceedings of the 2022 ACM SIGSAC Conference on Computer and Communications Security},
  pages={2595--2609},
  year={2022}
}

@inproceedings{duWindRangerDirectedGreybox2022,
  title={Windranger: A directed greybox fuzzer driven by deviation basic blocks},
  author={Du, Zhengjie and Li, Yuekang and Liu, Yang and Mao, Bing},
  booktitle={Proceedings of the 44th International Conference on Software Engineering},
  pages={2440--2451},
  year={2022}
}

@inproceedings{huangBEACONDirectedGreybox2022,
  title={Beacon: Directed grey-box fuzzing with provable path pruning},
  author={Huang, Heqing and Guo, Yiyuan and Shi, Qingkai and Yao, Peisen and Wu, Rongxin and Zhang, Charles},
  booktitle={2022 IEEE Symposium on Security and Privacy (SP)},
  pages={36--50},
  year={2022},
  organization={IEEE}
}

@inproceedings{srivastavaOneFuzzDoesn2022,
  title={One fuzz doesn’t fit all: Optimizing directed fuzzing via target-tailored program state restriction},
  author={Srivastava, Prashast and Nagy, Stefan and Hicks, Matthew and Bianchi, Antonio and Payer, Mathias},
  booktitle={Proceedings of the 38th Annual Computer Security Applications Conference},
  pages={388--399},
  year={2022}
}

@inproceedings{kimDAFLDirectedGreybox2023,
  title={{\{DAFL}\}: Directed Grey-box Fuzzing guided by Data Dependency},
  author={Kim, Tae Eun and Choi, Jaeseung and Heo, Kihong and Cha, Sang Kil},
  booktitle={32nd USENIX Security Symposium (USENIX Security 23)},
  pages={4931--4948},
  year={2023}
}

@article{wangEnhancingLargeLanguage2023,
  title={Enhancing Large Language Models for Secure Code Generation: A Dataset-driven Study on Vulnerability Mitigation},
  author={Wang, Jiexin and Cao, Liuwen and Luo, Xitong and Zhou, Zhiping and Xie, Jiayuan and Jatowt, Adam and Cai, Yi},
  journal={arXiv preprint arXiv:2310.16263},
  year={2023}
}

@inproceedings{heLargeLanguageModels2023,
  title={Large language models for code: Security hardening and adversarial testing},
  author={He, Jingxuan and Vechev, Martin},
  booktitle={Proceedings of the 2023 ACM SIGSAC Conference on Computer and Communications Security},
  pages={1865--1879},
  year={2023}
}

@article{de-fitero-dominguezEnhancedAutomatedCode2024,
  title={Enhanced Automated Code Vulnerability Repair using Large Language Models},
  author={de-Fitero-Dominguez, David and Garcia-Lopez, Eva and Garcia-Cabot, Antonio and Martinez-Herraiz, Jose-Javier},
  journal={arXiv preprint arXiv:2401.03741},
  year={2024}
}

@inproceedings{zhengCodeGeeXPretrainedModel2023,
  title={Codegeex: A pre-trained model for code generation with multilingual benchmarking on humaneval-x},
  author={Zheng, Qinkai and Xia, Xiao and Zou, Xu and Dong, Yuxiao and Wang, Shan and Xue, Yufei and Shen, Lei and Wang, Zihan and Wang, Andi and Li, Yang and others},
  booktitle={Proceedings of the 29th ACM SIGKDD Conference on Knowledge Discovery and Data Mining},
  pages={5673--5684},
  year={2023}
}

@inproceedings{wanImprovingAutomaticSource2018,
  title={Improving automatic source code summarization via deep reinforcement learning},
  author={Wan, Yao and Zhao, Zhou and Yang, Min and Xu, Guandong and Ying, Haochao and Wu, Jian and Yu, Philip S},
  booktitle={Proceedings of the 33rd ACM/IEEE international conference on automated software engineering},
  pages={397--407},
  year={2018}
}

@article{mathewsLLbezpekyLeveragingLarge2024,
  title={LLbezpeky: Leveraging Large Language Models for Vulnerability Detection},
  author={Mathews, Noble Saji and Brus, Yelizaveta and Aafer, Yousra and Nagappan, Mei and McIntosh, Shane},
  journal={arXiv preprint arXiv:2401.01269},
  year={2024}
}

@article{shestovFinetuningLargeLanguage2024,
  title={Finetuning Large Language Models for Vulnerability Detection},
  author={Shestov, Alexey and Cheshkov, Anton and Levichev, Rodion and Mussabayev, Ravil and Zadorozhny, Pavel and Maslov, Evgeny and Vadim, Chibirev and Bulychev, Egor},
  journal={arXiv preprint arXiv:2401.17010},
  year={2024}
}

@article{zhouLargeLanguageModel2024,
  title={Large Language Model for Vulnerability Detection: Emerging Results and Future Directions},
  author={Zhou, Xin and Zhang, Ting and Lo, David},
  journal={arXiv preprint arXiv:2401.15468},
  year={2024}
}

@article{nijkampCodeGenOpenLarge2022,
  title={Codegen: An open large language model for code with multi-turn program synthesis},
  author={Nijkamp, Erik and Pang, Bo and Hayashi, Hiroaki and Tu, Lifu and Wang, Huan and Zhou, Yingbo and Savarese, Silvio and Xiong, Caiming},
  journal={arXiv preprint arXiv:2203.13474},
  year={2022}
}

@article{friedInCoderGenerativeModel2022,
  title={Incoder: A generative model for code infilling and synthesis},
  author={Fried, Daniel and Aghajanyan, Armen and Lin, Jessy and Wang, Sida and Wallace, Eric and Shi, Freda and Zhong, Ruiqi and Yih, Wen-tau and Zettlemoyer, Luke and Lewis, Mike},
  journal={arXiv preprint arXiv:2204.05999},
  year={2022}
}

@article{cumminsLargeLanguageModels2023,
  title={Large language models for compiler optimization},
  author={Cummins, Chris and Seeker, Volker and Grubisic, Dejan and Elhoushi, Mostafa and Liang, Youwei and Roziere, Baptiste and Gehring, Jonas and Gloeckle, Fabian and Hazelwood, Kim and Synnaeve, Gabriel and others},
  journal={arXiv preprint arXiv:2309.07062},
  year={2023}
}

@article{zanLargeLanguageModels2023,
  title={Large language models meet nl2code: A survey},
  author={Zan, Daoguang and Chen, Bei and Zhang, Fengji and Lu, Dianjie and Wu, Bingchao and Guan, Bei and Wang, Yongji and Lou, Jian-Guang},
  journal={arXiv preprint arXiv:2212.09420},
  year={2022}
}

@inproceedings{devlinBERTPretrainingDeep,
  title={Bert: Pre-training of deep bidirectional transformers for language understanding},
  author={Kenton, Jacob Devlin Ming-Wei Chang and Toutanova, Lee Kristina},
  booktitle={Proceedings of naacL-HLT},
  volume={1},
  pages={2},
  year={2019}
}

@inproceedings{fuLineVulTransformerbasedLinelevel2022,
  title={Linevul: A transformer-based line-level vulnerability prediction},
  author={Fu, Michael and Tantithamthavorn, Chakkrit},
  booktitle={Proceedings of the 19th International Conference on Mining Software Repositories},
  pages={608--620},
  year={2022}
}

@inproceedings{fanCodeVulnerabilityDataset2020,
  title={AC/C++ code vulnerability dataset with code changes and CVE summaries},
  author={Fan, Jiahao and Li, Yi and Wang, Shaohua and Nguyen, Tien N},
  booktitle={Proceedings of the 17th International Conference on Mining Software Repositories},
  pages={508--512},
  year={2020}
}

@article{vaswaniAttentionAllYou2017,
  title={Attention is all you need},
  author={Vaswani, Ashish and Shazeer, Noam and Parmar, Niki and Uszkoreit, Jakob and Jones, Llion and Gomez, Aidan N and Kaiser, {\L}ukasz and Polosukhin, Illia},
  journal={Advances in neural information processing systems},
  volume={30},
  year={2017}
}

@inproceedings{choiSMARTIANEnhancingSmart2021,
  title={Smartian: Enhancing smart contract fuzzing with static and dynamic data-flow analyses},
  author={Choi, Jaeseung and Kim, Doyeon and Kim, Soomin and Grieco, Gustavo and Groce, Alex and Cha, Sang Kil},
  booktitle={2021 36th IEEE/ACM International Conference on Automated Software Engineering (ASE)},
  pages={227--239},
  year={2021},
  organization={IEEE}
}

@inproceedings{kimDriveFuzzDiscoveringAutonomous2022,
  title={Drivefuzz: Discovering autonomous driving bugs through driving quality-guided fuzzing},
  author={Kim, Seulbae and Liu, Major and Rhee, Junghwan" John" and Jeon, Yuseok and Kwon, Yonghwi and Kim, Chung Hwan},
  booktitle={Proceedings of the 2022 ACM SIGSAC Conference on Computer and Communications Security},
  pages={1753--1767},
  year={2022}
}

@inproceedings{liuStateSelectionAlgorithms2022,
  title={State selection algorithms and their impact on the performance of stateful network protocol fuzzing},
  author={Liu, Dongge and Pham, Van-Thuan and Ernst, Gidon and Murray, Toby and Rubinstein, Benjamin IP},
  booktitle={2022 IEEE International Conference on Software Analysis, Evolution and Reengineering (SANER)},
  pages={720--730},
  year={2022},
  organization={IEEE}
}

@inproceedings{hanCodeAlchemistSemanticsawareCode2019,
  title={CodeAlchemist: Semantics-aware code generation to find vulnerabilities in JavaScript engines.},
  author={Han, HyungSeok and Oh, DongHyeon and Cha, Sang Kil},
  booktitle={NDSS},
  year={2019}
}

@inproceedings{luoWestworldFuzzingassistedRemote2021,
  title={Westworld: Fuzzing-assisted remote dynamic symbolic execution of smart apps on iot cloud platforms},
  author={Luo, Lannan and Zeng, Qiang and Yang, Bokai and Zuo, Fei and Wang, Junzhe},
  booktitle={Proceedings of the 37th Annual Computer Security Applications Conference},
  pages={982--995},
  year={2021}
}

@misc{zalewski2021afl,
  author = {Michał Zalewski},
  title = {american fuzzy lop},
  year = {2021},
  url = {https://github.com/google/AFL},
  note = {Accessed: 2024-04-23}
}

@inproceedings{liSteelixProgramstateBased2017,
  title={Steelix: program-state based binary fuzzing},
  author={Li, Yuekang and Chen, Bihuan and Chandramohan, Mahinthan and Lin, Shang-Wei and Liu, Yang and Tiu, Alwen},
  booktitle={Proceedings of the 2017 11th joint meeting on foundations of software engineering},
  pages={627--637},
  year={2017}
}

@inproceedings{chenAngoraEfficientFuzzing2018,
  title={Angora: Efficient fuzzing by principled search},
  author={Chen, Peng and Chen, Hao},
  booktitle={2018 IEEE Symposium on Security and Privacy (SP)},
  pages={711--725},
  year={2018},
  organization={IEEE}
}

@inproceedings{lemieuxFairFuzzTargetedMutation2018,
  title={Fairfuzz: A targeted mutation strategy for increasing greybox fuzz testing coverage},
  author={Lemieux, Caroline and Sen, Koushik},
  booktitle={Proceedings of the 33rd ACM/IEEE international conference on automated software engineering},
  pages={475--485},
  year={2018}
}

@inproceedings{bohmeCoveragebasedGreyboxFuzzing2019,
  title={Coverage-based greybox fuzzing as markov chain},
  author={B{\"o}hme, Marcel and Pham, Van-Thuan and Roychoudhury, Abhik},
  booktitle={Proceedings of the 2016 ACM SIGSAC Conference on Computer and Communications Security},
  pages={1032--1043},
  year={2016}
}

@inproceedings{canakciTargetFuzzUsingDARTs2022,
  title={Targetfuzz: Using darts to guide directed greybox fuzzers},
  author={Canakci, Sadullah and Matyunin, Nikolay and Graffi, Kalman and Joshi, Ajay and Egele, Manuel},
  booktitle={Proceedings of the 2022 ACM on Asia conference on computer and communications security},
  pages={561--573},
  year={2022}
}

@inproceedings{luoSelectFuzzEfficientDirected2023,
  title={Selectfuzz: Efficient directed fuzzing with selective path exploration},
  author={Luo, Changhua and Meng, Wei and Li, Penghui},
  booktitle={2023 IEEE Symposium on Security and Privacy (SP)},
  pages={2693--2707},
  year={2023},
  organization={IEEE}
}

@inproceedings{osterlundParmeSanSanitizerguidedGreybox2020,
  title={{\{ParmeSan}\}: Sanitizer-guided greybox fuzzing},
  author={{\"O}sterlund, Sebastian and Razavi, Kaveh and Bos, Herbert and Giuffrida, Cristiano},
  booktitle={29th USENIX Security Symposium (USENIX Security 20)},
  pages={2289--2306},
  year={2020}
}

@inproceedings{zongFuzzGuardFilteringOut2020,
  title={{\{FuzzGuard}\}: Filtering out unreachable inputs in directed grey-box fuzzing through deep learning},
  author={Zong, Peiyuan and Lv, Tao and Wang, Dawei and Deng, Zizhuang and Liang, Ruigang and Chen, Kai},
  booktitle={29th USENIX security symposium (USENIX security 20)},
  pages={2255--2269},
  year={2020}
}

@inproceedings{maDirectedSymbolicExecution2011,
  title={Directed symbolic execution},
  author={Ma, Kin-Keung and Yit Phang, Khoo and Foster, Jeffrey S and Hicks, Michael},
  booktitle={Static Analysis: 18th International Symposium, SAS 2011, Venice, Italy, September 14-16, 2011. Proceedings 18},
  pages={95--111},
  year={2011},
  organization={Springer}
}

@inproceedings{marinescuKATCHHighcoverageTesting2013,
  title={KATCH: High-coverage testing of software patches},
  author={Marinescu, Paul Dan and Cadar, Cristian},
  booktitle={Proceedings of the 2013 9th Joint Meeting on Foundations of Software Engineering},
  pages={235--245},
  year={2013}
}

@article{yangDirectedIncrementalSymbolic2014,
  title={Directed incremental symbolic execution},
  author={Person, Suzette and Yang, Guowei and Rungta, Neha and Khurshid, Sarfraz},
  journal={Acm Sigplan Notices},
  volume={46},
  number={6},
  pages={504--515},
  year={2011},
  publisher={ACM New York, NY, USA}
}

@inproceedings{christakisGuidingDynamicSymbolic2016,
  title={Guiding dynamic symbolic execution toward unverified program executions},
  author={Christakis, Maria and M{\"u}ller, Peter and W{\"u}stholz, Valentin},
  booktitle={Proceedings of the 38th International Conference on Software Engineering},
  pages={144--155},
  year={2016}
}

@inproceedings{cadarKLEEUnassistedAutomatic2008,
  title={Klee: unassisted and automatic generation of high-coverage tests for complex systems programs.},
  author={Cadar, Cristian and Dunbar, Daniel and Engler, Dawson R and others},
  booktitle={OSDI},
  volume={8},
  pages={209--224},
  year={2008}
}

@article{wang2023syztrust,
  title={SyzTrust: State-aware Fuzzing on Trusted OS Designed for IoT Devices},
  author={Wang, Qinying and Chang, Boyu and Ji, Shouling and Tian, Yuan and Zhang, Xuhong and Zhao, Binbin and Pan, Gaoning and Lyu, Chenyang and Payer, Mathias and Wang, Wenhai and others},
  journal={arXiv preprint arXiv:2309.14742},
  year={2023}
}

@inproceedings{huang2024everything,
  title={Everything is Good for Something: Counterexample-Guided Directed Fuzzing via Likely Invariant Inference},
  author={Huang, Heqing and Zhou, Anshunkang and Payer, Mathias and Zhang, Charles},
  booktitle={2024 IEEE Symposium on Security and Privacy (SP)},
  pages={142--142},
  year={2024},
  organization={IEEE Computer Society}
}

\end{document}